\documentclass{emulateapj}
\usepackage{apjfonts}

\newcommand{\Msun}      {\mbox{$\rm\,M_{\mathord\odot}$}}

\submitted{Accepted by the Astrophysical Journal}

\begin{document}

\def\lsim{\mathrel{\lower .85ex\hbox{\rlap{$\sim$}\raise
.95ex\hbox{$<$} }}}
\def\gsim{\mathrel{\lower .80ex\hbox{\rlap{$\sim$}\raise
.90ex\hbox{$>$} }}}

\lefthead{Two Years of X-Ray Activity From 4U 1630--47}
\righthead{J.A. Tomsick et al.}

\title{X-Ray Observations of the Black Hole Transient 4U~1630--47 
During Two Years of X-Ray Activity}

\author{John A. Tomsick\altaffilmark{1},
St\'ephane Corbel\altaffilmark{2},
Andrea Goldwurm\altaffilmark{3},
Philip Kaaret\altaffilmark{4}}

\altaffiltext{1}{Center for Astrophysics and Space Sciences, Code
0424, University of California at San Diego, La Jolla, CA,
92093, USA (e-mail: jtomsick@ucsd.edu)}

\altaffiltext{2}{Universit\'e Paris VII and Service d'Astrophysique,
CEA Saclay, 91191 Gif sur Yvette, France}

\altaffiltext{3}{Service d'Astrophysique, CEA Saclay, 91191 Gif 
sur Yvette and APC, 11 place M. Berthelot, 75231 Paris, France}

\altaffiltext{4}{Department of Physics and Astronomy, University of
Iowa, Iowa City, IA 52242, USA}

\begin{abstract}

The black hole candidate (BHC) X-ray transient 4U~1630--47 
continuously produced strong X-ray emission for over 2 years 
during its 2002--2004 outburst, which is one of the brightest 
and longest outbursts ever seen from this source.  We use 
over 300 observations made with the {\em Rossi X-ray Timing 
Explorer (RXTE)} to study the source throughout the outburst
along with hard X-ray images from the {\em INTErnational 
Gamma-Ray Astrophysics Laboratory}, which are critical for 
interpreting the {\em RXTE} data in this crowded field.  
The source exhibits extreme behaviors, which can be interpreted
as an indication that the system luminosity approaches the 
Eddington limit.  For 15 observations, fitting the spectral 
continuum with a disk-blackbody plus power-law model results 
in measured inner disk temperatures between 2.7 and 3.8~keV, 
and such temperatures are only rivaled by the brightest BHC 
systems such as GRS~1915+105 and XTE~J1550--564.  If the high 
temperatures are caused by the dominance of electron scattering 
opacity in the inner regions of the accretion disk, it is 
theoretically required that the source luminosity is considerably 
higher than 20\% of the Eddington limit.  We detect a variety
of high-amplitude variability, including hard 10--100~s flares, 
which peak at levels as much as 2--3 times higher than non-flare 
levels.  This flaring occurs at the highest disk luminosities in 
a regime where the source deviates from the 
$L_{\rm disk}\propto T_{\rm in}^{4}$ relationship that is seen
at lower luminosities, possibly suggesting that we are seeing 
transitions between a Shakura \& Sunyaev disk and a ``slim'' 
disk, which is predicted to occur at very high mass accretion 
rates.  The X-ray properties in 2002--2004 are significantly 
different from those seen during the 1998 outburst, which is 
the only outburst with detected radio jet emission.  Our results 
support the ``jet-line'' concept recently advanced by Fender 
and co-workers.  Our study allows for a test of the 
quantitative McClintock \& Remillard spectral state definitions, 
and we find that these definitions alone do not provide a 
complete description of the outburst.  Finally, for several 
of the observations, the high energy emission is dominated 
by the nearby sources IGR J16320--4751 and IGR J16358--4726,
and we provide information on when these sources were 
bright and on the nature of their energy spectra.

\end{abstract}


\section{Introduction}

The X-ray ($\sim$1--10 keV) luminosities from Galactic black hole 
candidate (BHC) transients range from values below $10^{30}$ erg~s$^{-1}$ 
when the sources are in quiescence \citep{garcia01} to values that can 
approach or exceed $10^{39}$ erg~s$^{-1}$ for some sources \citep{dwg04}.  
During outbursts, luminosities above $\sim$$10^{34}$ erg~s$^{-1}$ 
are usually seen for at least several months \citep{csl97}, and 
significant changes in the X-ray properties occur over time.  We 
do not have a detailed understanding of all the physical changes that 
lead to changes in the X-ray emission properties, but the physics 
involves the structure of the accretion disk around the black hole 
as well as the connection between the accretion disk and the steady 
or impulsive jets that can be launched from these systems.  The 
changes in the X-ray emission properties are partially caused by 
changes in the mass accretion rate onto the black hole; however, it 
has been demonstrated that other physical parameters must also be 
important for determining those properties \citep{homan00,tomsick04_rossi}.

The emission properties of accreting black holes are often classified 
in terms of ``spectral states.''  Recently, efforts have been made 
to make the state definitions more quantitative and to connect these
definitions directly to the continuum spectral components \citep{mr03}.
The spectra can often be described as the combination of a soft, 
thermal component along with a hard component that can fall off more
or less steeply with energy.  The thermal component is almost certainly
blackbody emission from an accretion disk, as in \cite{ss73}, but the
mechanism for producing the hard component is less clear.  Accreting
black holes can be highly variable, sometimes with quasi-periodic 
oscillations (QPOs); spectral and timing properties are both
incorporated in the following spectral state definitions from
\cite{mr03}.

In the Thermal-Dominant (TD) state, the thermal component accounts 
for $>$75\% of the total 2--20~keV flux.  In this state, no or weak 
QPOs are seen with rms levels below 1\%, and the 0.1--10~Hz continuum 
rms is $\lsim$6\%.  In the Steep Power-Law (SPL) state, the hard 
component is a power-law with $\Gamma > 2.4$, where $\Gamma$ is the 
power-law photon index.  A source is said to be in the SPL state if 
QPOs are present and the hard component contributes $>$20\% of the 
2--20~keV flux or if, regardless of the timing properties, the hard 
component contributes $> 50$\% of the 2--20~keV flux.  Finally, in 
the Hard state, the hard component is much less steep, at 
$1.5 < \Gamma < 2.1$, the hard component contributes more than 80\% 
of the 2--20~keV flux, the 0.1--10~Hz continuum rms is between 10\% 
and 30\%, and the presence of radio emission signals the presence of 
a compact jet \citep{fender01}.  As discussed in \cite{mr03}, systems 
also exhibit intermediate states, with properties that are usually 
some combination of the three main states (TD, SPL, and Hard).

The X-ray activity from the BHC transient 4U~1630--47 over more
than two years during 2002--2004 along with good high energy coverage 
of the source by the {\em Rossi X-ray Timing Explorer (RXTE)} and 
the {\em INTErnational Gamma-Ray Astrophysics Laboratory (INTEGRAL)} 
provide an opportunity to study the long-term evolution of the
source as it enters different spectral states and exhibits 
different emission properties.  4U~1630--47 is among the most active 
of the BHC transients and has produced strong hard X-ray emission 
during its 17 detected outbursts 
\citep{tbp01,tk00,oosterbroek98,kuulkers97,parmar97}.  The source 
has a quasi-periodic $\sim$600--700~day outburst recurrence time 
\citep{kuulkers97}, which is unusually short for systems of this 
type, probably indicating a higher time-averaged mass accretion 
rate from its companion \citep{csl97}.  Highly polarized radio
emission was detected from 4U~1630--47 during its 1998 outburst
\citep{hjellming99}, indicating the presence of jets, and the 
source is often compared to microquasars such as GRS~1915+105 and 
GRO~J1655--40.  The compact object mass has not been measured for 
4U~1630--47, but \cite{mr03} classify it as a very likely 
``category A'' BHC.  Also, the binary orbital period is not known, 
and our lack of knowledge is due to the difficulty in performing 
optical and infrared studies of the source due to its high column 
density \citep[but see][for the likely identification of the 
source's infrared counterpart]{akv01}.

\begin{figure}
\centerline{\includegraphics[width=0.52\textwidth]{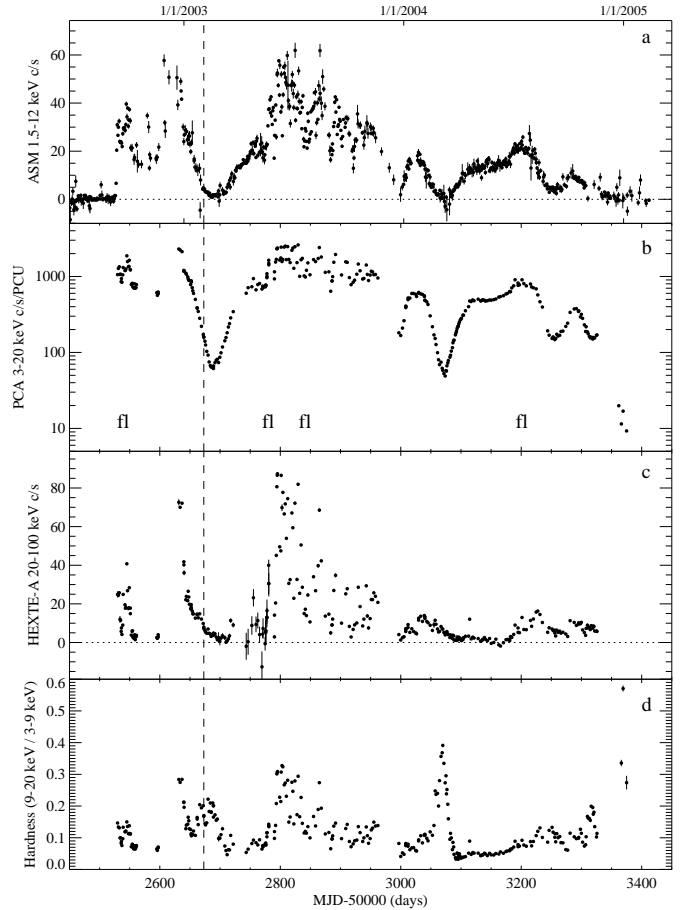}}
\caption{X-ray light curves and hardness vs. time for
4U~1630--47 from 2002 September to 2005 January.  ({\em a}) shows 
the {\em RXTE} ASM 1.5--12 keV rates (daily averages). ({\em b}) 
shows the 3--20 keV PCA light curve for the 318 pointed observations.  
({\em c}) shows the 20--100 keV HEXTE light curve.  ({\em d}) shows 
the 9--20/3--9 keV hardness ratio using the PCA count rates.  The 
vertical dashed line marks the time of the {\em INTEGRAL} observation.  
In panel {\em b}, ``fl'' marks the examples of the flaring behavior 
shown in Figure~\ref{fig:lc16f}.  Also, the PCA count rates for the 
final four observations (see panel {\em b}) are consistent with little 
or no contribution to the emission from 4U~1630--47.\label{fig:lc}}
\end{figure}

The current outburst from 4U~1630--47, which began in 2002 
September \citep{wrm02}, is one of the brightest and longest 
recorded outbursts from this system.  A new high-amplitude
flaring behavior has been reported at different times during
this outburst \citep{hw02,tomsick03_atel}, and, in this work, 
we study this behavior in more detail.  Also, transitions 
between spectral states have been observed during this outburst
\citep{tomsick04_munich,tomsick04_atel}.  In addition to the high 
level of recent activity from 4U~1630--47 and the recent work 
on defining spectral states, our study comes at a time when 
{\em INTEGRAL} is providing high quality hard X-ray images.  
Images of the 4U~1630--47 field, which we present in this work 
and in \cite{tomsick04_munich}, are extremely useful for avoiding
source confusion.  In the following, we present X-ray spectral
and timing studies of 4U~1630--47. 

\section{Observations}

{\em RXTE} regularly monitored 4U~1630--47 in outburst with pointed 
observations between 2002 September 12 (MJD 52,529) and 2005 January 4 
(MJD 53,375).  Here, we study the evolution of the X-ray properties 
throughout the outburst by analyzing data from the 318 {\em RXTE} 
observations that occurred during this time.  With the exception of 
five 22-36 day observing gaps caused mainly by sun angle constraints, 
pointed observations occurred, on average, approximately every other 
day for 2.3 years.  The observations were made under seven different 
proposal IDs (see Table~\ref{tab:observations}): In three cases 
(P70113, P80117, and P90128), we observed 4U~1630--47 in conjunction 
with our {\em INTEGRAL} program; and, in the other four cases (P70417, 
P80417, P80420, and P90410), the observations were made under a public 
``Target of Opportunity'' program, and we analyzed data from the public 
archive.  The Proportional Counter Array (PCA) and High Energy X-ray 
Timing Experiment (HEXTE) light curves shown in Figure~\ref{fig:lc} 
indicate the times of the {\em RXTE} observations.

After the launch of the ESA satellite {\em INTEGRAL} \citep{winkler03} 
in 2002 October, we triggered our Target of Opportunity to observe 
4U~1630--47 when the source became observable while it was in
outburst in early 2003.  We obtained a 293 ks exposure between
UT 2003 February 1.2 and UT 2003 February 5.3, and the time of
this observation is marked in Figure~\ref{fig:lc}.  We previously 
used the data from this observation for studies of the obscured 
X-ray source IGR J16320--4751 (= AX J1631.9--4752), which is a 
hard and persistent (though highly variable) source that is 
close to 4U~1630--47 \citep{tomsick03_iauc,rodriguez03,foschini04}, 
and we used these data to report detections of two other new 
{\em INTEGRAL} sources \citep{tomsick04_new}.  In addition, the 
20--40 keV images from this observation can be found in 
\cite{tomsick04_munich} along with information about the 
sources in the {\em INTEGRAL} field of view, including an 
initial look at the hard X-ray and gamma-ray properties of
4U~1630--47.  Although we do not present a full analysis of
the {\em INTEGRAL} data here, the hard X-ray {\em INTEGRAL} images 
are important for interpreting the {\em RXTE} data.

\begin{figure}
\centerline{\includegraphics[width=0.48\textwidth]{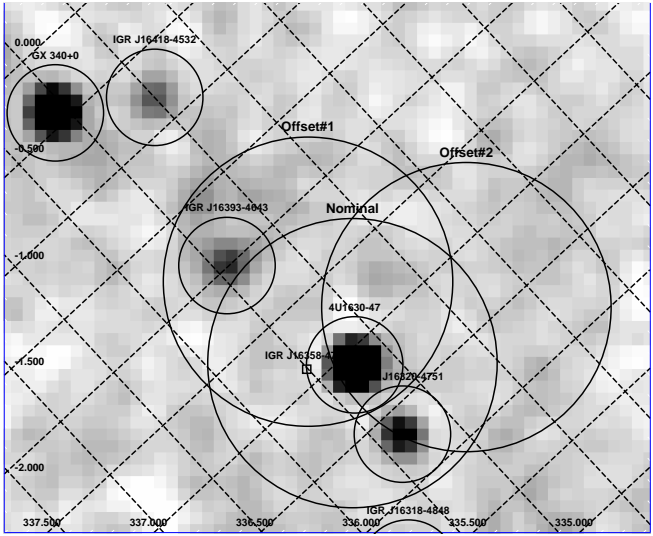}}
\caption{A 20--40 keV image (significance map) using 
the IBIS instrument on {\em INTEGRAL} taken on UT 2003 February 
1.2--5.3.  The small circles mark detected sources, and the position 
of the undetected transient IGR J16358--4726 is marked with a square.  
The large circles, labeled ``Nominal,'' ``Offset\#1,'' and ``Offset\#2,'' 
show the three pointing positions used during the {\em RXTE} campaign.  
The circles have a $2^{\circ}$ diameter, corresponding to the 
full-width zero-intensity (FWZI) collimator response.  We used
different {\em RXTE} pointing positions in an attempt to minimize 
contamination of 4U~1630--47 by nearby sources.\label{fig:image}}
\end{figure}

\section{Data Analysis}

We extracted energy spectra and light curves from the {\em RXTE} 
data using scripts developed at the University of California,
San Diego and the University of T\"{u}bingen that incorporate 
the standard software for {\em RXTE} data reduction (FTOOLS).
We processed the data using the most recent {\em RXTE} 
calibration files, which were released on 2003 July 7, and 
performed extractions using FTOOLS v5.3 and v5.3.1.  The
{\em RXTE} routines are identical for these two versions
of FTOOLS.  We performed time filtering for the 330 
observations using data from times during which the following 
criteria are satisfied:  The {\em RXTE} pointing is within 
$0^{\circ}.1$ of the nominal pointing position; the nominal 
pointing position is more than $10^{\circ}$ above the limb of 
the Earth; a South Atlantic Anomaly (SAA) passage has not occurred 
within the previous 30 minutes; Proportional Counter Unit
(PCU) 2 is turned on; and the PCU 2 electron ratio is less
than 0.25.  We chose PCU 2 because, in normal {\em RXTE}
operation, this unit is programmed to have a very high duty 
cycle to allow for the most precise observation-to-observation
comparisons.  Our filtering led to zero exposure time for only
12 out of 330 observations (3.6\%), leaving us with 318
observations for further study.  As shown in Table~\ref{tab:observations},
we typically obtained 1--3 ks of exposure time per observation, 
but exposure times vary greatly from 96 s up to 11 ks.
For all 318 observations, the mean exposure time is 1.8 ks.

We extracted the following information for each observation:
The 3--20 keV PCU 2 count rate; the 3--9 keV PCU 2 count rate;
the 9--20 keV PCU 2 count rate; the 20--100 keV HEXTE-A count
rate; the 3--200 keV PCA+HEXTE energy spectrum; and the 
3--20 keV PCA light curve with 16 s time bins.  We took the
PCA information from the ``Standard 2'' data, which includes
129 channel energy spectra taken with 16 s time resolution.
We used the ``Sky-VLE'' model to estimate and subtract off
the background.  For HEXTE, we used ``Event List'' data, 
and we used the normal HEXTE rocking mode to estimate and
subtract off the background.  As described below in \S$7$, 
we also selected five observations for more detailed study, 
and, for these, we used the higher time resolution PCA data 
to produce power spectra.  For four of the observations, we
used a PCA mode with 64 energy channels and $2^{-13}$~s
(= 122 $\mu$s) time resolution.  In the fifth case, we
combined the data from two PCA modes:  A mode with 
$2^{-7}$~s (7.8 ms) time resolution, covering the lower
energy portion of the spectrum; and a second mode with 
higher time resolution, containing an event list for the
higher energy photons.  We used data from all the active
PCUs when producing the power spectra.

{\em INTEGRAL} hard X-ray images from the observation 
described above, show that 4U~1630--47 lies in a region 
of the Galaxy that has a high density of hard X-ray 
sources, including a source, IGR J16320--4751, that is
$0^{\circ}.58$ from 4U~1630--47.  The 20--40 keV Imager
on-Board the {\em INTEGRAL} Satellite (IBIS) image, produced 
using the Off-line Scientific Analysis (OSA-4.2) software
\citep{goldwurm03}, is shown in Figure~\ref{fig:image}.  When 
we realized that there is another hard X-ray source within the 
{\em RXTE} field-of-view (FOV) for the nominal 4U~1630--47 
pointing position, we requested a change in the pointing 
position to avoid IGR J16320--4751.  From MJD 52,691 to 
52,722, we used the pointing position marked ``Offset \#1'' 
in Figure~\ref{fig:image}.  However, the presence of 
IGR J16393--4643 \citep{combi04} and a third X-ray transient, 
IGR J16358--4726 \citep{patel04}, prompted another change in 
pointing position to ``Offset \#2,'' and we used this 
pointing position from MJD 52,722 to 52,781.  The PCA and
HEXTE collimators have a triangular response with a 
full-width half-maximum (FWHM) FOV of $1^{\circ}$ and 
a full-width zero-intensity (FWZI) FOV of $2^{\circ}$.  
Where necessary in the work described below, we corrected 
for the collimator response when producing light curves or 
response matrices for spectral analysis.

\section{Flux and Hardness Evolution During the Outburst}

Figure~\ref{fig:lc}a shows the {\em RXTE} All-Sky Monitor 
(ASM) 1.5--12 keV light curve for 4U~1630--47 with the source
in outburst from 2002 September to the end of 2004.  In 
addition to being the longest outburst during the {\em RXTE} 
lifetime, at its peak, its ASM flux is $\sim$800 mCrab 
(1 Crab = 74 ASM c/s), which is $\sim$50\% brighter than 
any previous outburst observed by the ASM.  Below, we divide 
the outburst into the various spectral states the source 
entered, but, in general, the ASM light curve shows two very 
bright and highly variable periods:  The first occurred for 
$\sim$120 days between MJD 52,530 and $\sim$52,650; and the 
second occurred for $\sim$200 days in 2003 between MJD 52,750 
and 52,950.  The source was also bright for much of 2004, 
but it did not become as bright as the previous two periods
of very high activity.  Twice in 2003 and once in early 2004, 
the source flux was low enough to be only marginally detected 
by the ASM.  The source became undetectable again in late 
2004, and it appears that the source has remained in
quiescence into 2005.  

Figure~\ref{fig:lc}b shows the count rates measured in the 
3--20 keV band by the PCA during the pointed {\em RXTE} 
observations.  For the observations made at the offset 
pointing positions, the count rates are corrected using the 
PCA collimator response.  The {\em RXTE} monitoring program 
began soon after the source was detected by the ASM.  
While the source flux was at the ASM sensitivity limit at 
times, the PCA, with its better sensitivity, shows continuous 
activity until 2004 November 16 (MJD 53,326).  After this 
date, there was a 36 day gap in coverage, and we obtained 
four more PCA measurements after this gap.  For these 
observations, the PCA rates are consistent with Galactic 
ridge emission and flux from IGR~J16320--4751, with little 
or no emission from 4U~1630--47.  They are not included 
in the spectral and timing analysis in the next two sections
(leaving 314 observations).  Figure~\ref{fig:lc}c shows the 
HEXTE-A 20--100 keV count rate, and Figure~\ref{fig:lc}d shows 
the source hardness, defined as the ratio of the 9--20 keV PCA 
rate to the 3--9 keV PCA rate.

\section{Energy Spectra}

For the 314 observations, we used the XSPEC v11.3.1t software to 
perform $\chi^{2}$-minimization spectral fits to the PCA+HEXTE
3--200 keV energy spectra.  For the PCA spectra, we included 
systematic errors at a level of 0.6\% for 3--8 keV and at a level 
of 0.3\% for 8--25 keV, and these numbers are derived by fitting
energy spectra of the Crab nebula as described in \cite{tck01}.
For many of the observations, the standard two component model -- 
disk-blackbody \citep{makishima86} plus power-law with interstellar 
absorption -- provides acceptable fits, but, for 31 observations, 
we obtain $\chi^{2}_{\nu} > 2.0$ for 63 degrees of freedom (dof), 
and, for the 314 observations, the mean $\chi^{2}_{\nu}$ is 1.51.  
By examining the residuals for several of the spectra with 
statistically poor fits, we found two main reasons for the poor 
fits.  First, in many cases, large residuals (positive and 
negative) are present around the iron K$\alpha$ complex.  Second, 
negative residuals are sometimes seen at high energies, above 
50 keV, indicating the presence of a cutoff in the spectrum.  The 
presence of iron features and high energy cutoffs are not 
surprising as they have been seen previously for 4U~1630--47 as 
well as for other black hole sources \citep{tk00,zdziarski96}.  

Thus, we refitted all the energy spectra after adding a 
narrow iron K$\alpha$ emission line and a smeared iron edge
\citep{ebisawa94} in a similar manner to that described in
\cite{tk00}.  We restricted the line energies to be between
6.4 keV and 7.1 keV, spanning the possible iron K$\alpha$
range for non-redshifted lines.  Similarly, we restricted
the edge energy to be between 7.1 keV and 9.3 keV.  Sample
fits to several spectra show that the width of the smeared
edge is not well-constrained in most cases, and we fixed 
the width to 10 keV \citep{ebisawa94,tk00}.  We also added 
a high energy cutoff allowing the model to exponentially 
turn-over above an energy $E_{\rm cut}$ with an e-folding 
energy of $E_{\rm fold}$.  Finally, although the column
density ($N_{\rm H}$) has been fixed to values close to
$10^{23}$ cm$^{-2}$ in some previous studies 
\citep[e.g.,][]{tk00}, the results for this {\em RXTE}
data set indicate significant changes during the outburst.  
Thus, we have left $N_{\rm H}$ as a free parameter, except 
that we have restricted the column density to be greater 
than $6\times 10^{22}$ cm$^{-2}$, which is the lowest value 
that has been measured for 4U~1630--47 by a soft X-ray 
instrument \citep{parmar97}.  This value may represent the 
interstellar value along the line of sight.

\begin{figure}
\centerline{\includegraphics[width=0.52\textwidth]{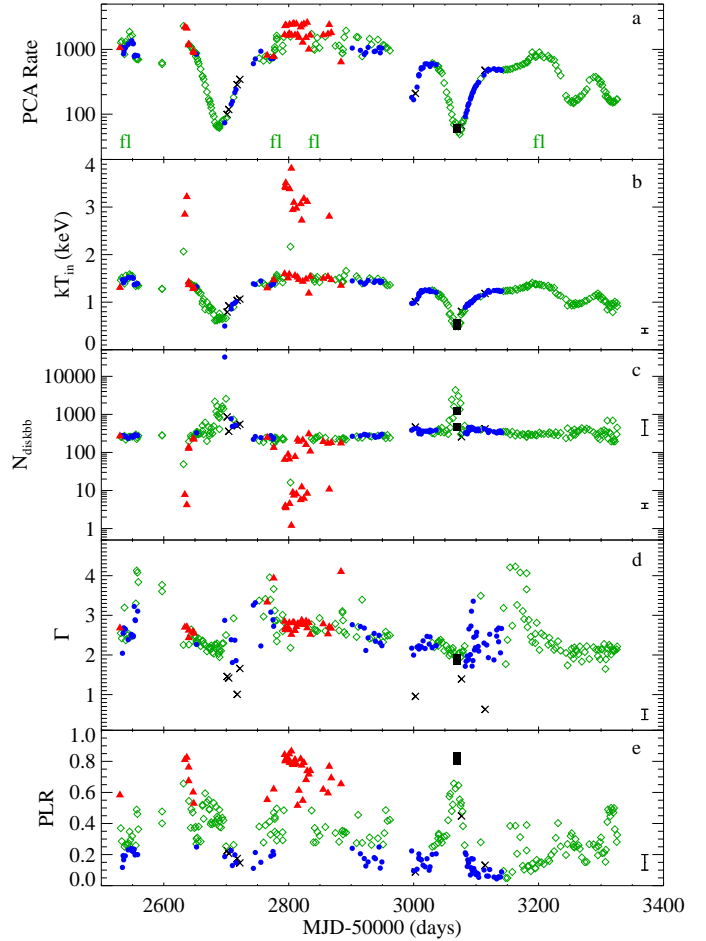}}
\vspace{0.5cm}
\caption{The evolution of the spectral properties 
of 4U~1630--47 during the 2002--2004 outburst.  The 314 observations
have been divided into spectral states using the spectral parameters
and the quantitative definitions of \cite{mr03}.  The colors/symbols 
correspond to states as follows:  Red triangles = Steep Power-Law; 
Green open diamonds = Intermediate State; Blue circles = Thermal-Dominant; 
Black squares = Hard.  The crosses indicate observations that we believe 
are contaminated by the nearby sources IGR~J16320--4751 or 
IGR J16358--4726.  ({\em a}) shows the 3--20~keV PCA light curve, and 
``fl'' marks the examples of the flaring behavior shown in 
Figure~\ref{fig:lc16f}.  ({\em b}) and ({\em c}) show the disk-blackbody 
parameters.  ({\em d}) shows the power-law photon index.  ({\em e}) shows 
the power-law ratio (see text for definition).  In panels ({\em b}) 
through ({\em e}), an average-sized error bar is shown, although it 
should be noted that there is some variation in the error bar size from 
observation-to-observation (see Table~\ref{tab:spectra}).  
\label{fig:parameters}}
\end{figure}

These additions to the spectral model produce significant
improvements in the quality of the fits.  For the 314
spectra, the mean $\chi^{2}_{\nu}$ is 1.15 for 57 dof, and
only 13 have $\chi^{2}_{\nu} > 2.0$.  We examined the 
spectral residuals as well as the 16 s light curves for 
the cases where the worst fits are obtained.  In most of
these cases, the light curves show a high degree of variability, 
suggesting that spectral variability during the observation 
degrades the quality of the fits.  However, in several cases, 
the level of variability is not particularly high, and, in 
four of these observations, significant positive residuals are 
present at high energies, above $\sim$40 keV.  Although it is 
possible that these residuals indicate a high energy excess 
from 4U~1630--47, the observations for which the excess is 
present occurred at the nominal pointing position for 4U~1630--47 
so that the {\em RXTE} field of view includes the persistent 
hard X-ray source IGR J16320--4751 (cf. Figure~\ref{fig:image}).  
Using {\em INTEGRAL} flux measurements and {\em RXTE} pointings 
at the 4U~1630--47 position during times when 4U~1630--47 was 
not active, we demonstrate in Appendix A that the positive high 
energy residuals are very likely caused by IGR J16320--4751.  
The four observations with positive high energy residuals 
occurred between MJD 52,560 and 52,596, and, as the spectra for 
these four observations are likely contaminated, we do not 
consider them in the following analysis.

Figure~\ref{fig:parameters} shows the PCA rates and spectral 
parameters vs.~time.  The spectral parameters shown include the 
temperature of the disk-blackbody component ($kT_{\rm in}$), 
the normalization of this component ($N_{\rm diskbb} = 
(R_{\rm in}/d_{10})^2\cos i$, where $R_{\rm in}$ is the inner 
radius of the accretion disk in km, $d_{10}$ is the source 
distance in units of 10 kpc, and $i$ is the disk inclination), 
the power-law photon index ($\Gamma$), and the ratio of the
unabsorbed 2--20~keV power-law flux to the total flux ($PLR$).
When the entire outburst is considered, the range of $kT_{\rm in}$ 
values is 0.49--3.81 keV, while the range is 0.63--4.13 for $\Gamma$.  
We use the spectral parameters and the \cite{mr03} state definitions 
to divide the observations into different spectral states, and the 
classifications are indicated in Figure~\ref{fig:parameters}.  The 
Steep Power-Law (SPL) observations have $\Gamma > 2.4$ and $PLR > 0.5$.  
Those in the Thermal-Dominant (TD) state have $PLR < 0.25$.  The 
source was only in the Hard state with $1.5 < \Gamma < 2.1$ and 
$PLR > 0.80$ for two observations near MJD 53,069.  The remainder 
of the observations do not fit into any of the spectral states
as defined by \cite{mr03}, and we say that the source was in
one or more intermediate states (IS) during these observations.  
It should be noted that we have divided the observations into 
spectral states using only spectral and not timing information.  
While timing information is important to the study of spectral 
states and is part of the full \cite{mr03} definitions, our 
results for a sample of the power spectra (see \S$7$) show that 
the timing properties are in-line with the \cite{mr03} criteria 
for the various states.  If we had included timing information 
in dividing up the spectral states, it is possible that some TD 
observations with higher noise levels would be re-classified as 
IS and that if some of the IS or TD observations have QPOs they 
may be re-classified as SPL.  However, such re-classifications 
would not change the results or conclusions of this work.

There are seven observations with distinctly different spectra.
For these observations, the power-law is very hard with values 
of $\Gamma$ between 0.6 and 1.7, and it is sharply cutoff with 
e-folding energies between 10 and 50 keV.  While such a spectrum 
would represent the discovery of a new black hole state, we 
strongly suspect that these observations are contaminated by 
emission from the hard X-ray transient IGR~J16358--4726.  Although 
IGR~J16358--4726 was not detected during our 2003 February 
{\em INTEGRAL} observation (see Figure~\ref{fig:image}), four of 
the observations for which we see the very hard X-ray spectrum 
occurred during the time period from MJD 52,701 to 52,722, when the 
source was known to be active 
\citep{revnivtsev03a,revnivtsev03b,patel04}.  During this time
period, we used the Offset \#1 pointing position, so the field 
of view included IGR~J16358--4726 but not IGR~J16320--4751.  In 
addition, the hard spectrum is not detected for any of the 
observations for which we used the Offset \#2 pointing position, 
which does not include IGR~J16358--4726 in the field of view.  
In Appendix B, we include details about the hard spectra, as 
they provide useful information on IGR~J16358--4726.  Other than 
in Appendix B, we do not consider these observations further.

\section{High-Amplitude Variability}

An inspection of the 3--20 keV 16~s PCA light curves for
all 314 observations indicates that in addition to the 
observation-to-observation variability that is clearly 
seen in Figures~\ref{fig:lc} and \ref{fig:parameters}, 
many of the light curves show high amplitude variability 
during the observations.  To quantify the level of 
variability, we calculated the peak-to-peak amplitude 
for each observation, $A_{\rm pp}$, defined simply as 
the maximum PCA count rate minus the minimum rate divided 
by the mean rate for the observation.  The errors on 
$A_{\rm pp}$ depend on the uncertainties in the count 
rates for the individual maximum and minimum 16 s time 
bins as well as the error in the mean rate.
For the 314 observations, the mean value of $A_{\rm pp}$ 
is 0.26 and the standard deviation is 0.20.  While the 
majority of the observations have some form of significant 
variability, we focus on the observations with high-amplitude 
variability defined as the observations during which
$A_{\rm pp}$ minus the 2-$\sigma$ error on $A_{\rm pp}$ is 
greater than 0.3.  With this definition, there are 72
observations with high-amplitude variability.  

\begin{figure}
\centerline{\includegraphics[width=0.52\textwidth]{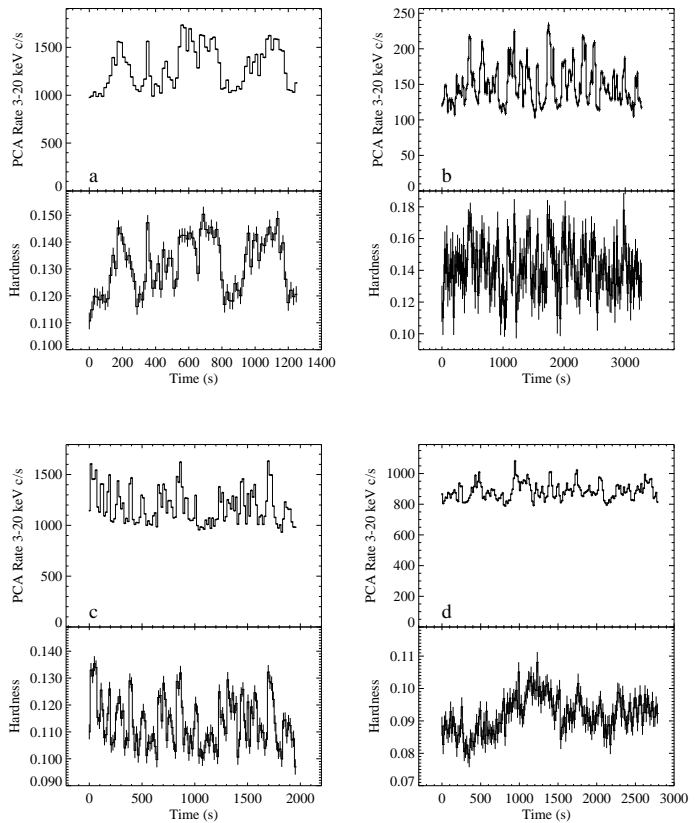}}
\caption{PCA light curves and 9--20/3--9 keV hardnesses 
vs. time with 16~s time bins for four of IS observations that exhibit
the flaring behavior.  The times and Observation IDs for these observations 
are ({\em a}) MJD 52,539.094 and 70417-01-04-02, ({\em b}) MJD 52,780.746
and 70113-01-43-00, ({\em c}) MJD 52,843.379 and 80117-01-13-02, and
({\em d}) MJD 53,201.305 and 90410-01-13-01.  These are also marked on
Figures~\ref{fig:lc} and \ref{fig:parameters}.  Note that the observations 
for ({\em a}), ({\em c}), and ({\em d}) were made in the nominal pointing 
position while the observation for ({\em b}) was made in one of the offset 
pointing positions.\label{fig:lc16f}}
\end{figure}

\begin{figure}
\centerline{\includegraphics[width=0.52\textwidth]{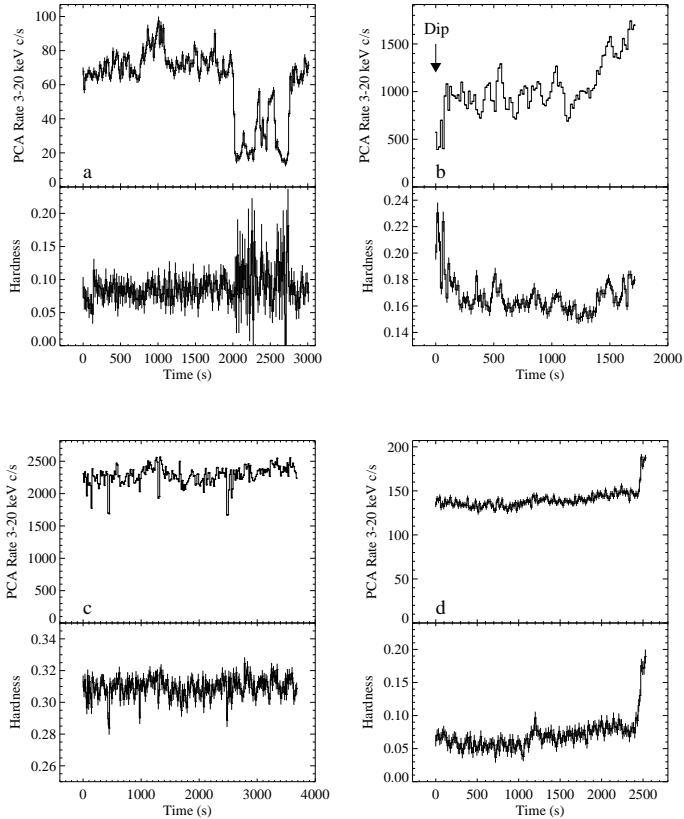}}
\caption{PCA light curves and 9--20/3--9 keV hardnesses 
vs. time with 16~s time bins showing three other types of high-amplitude
variability.  ({\em a}) and ({\em b}) show examples where absorption is
likely the cause of the dips.  ({\em c}) shows an example of short dips
seen in three of the SPL observations.  ({\em d}) shows an unusually hard
flare that is probably from 4U~1630--47.  The times and Observation IDs 
for these observations are ({\em a}) MJD 52,759.672 and 70113-01-36-00, 
({\em b}) MJD 52,795.340 and 80117-01-03-00, ({\em c}) MJD 52,832.188
and 80117-01-12-00, and ({\em d}) MJD 53,087.832 and 90128-01-05-00.  
Note that the observations for ({\em b}), ({\em c}), and ({\em d}) were 
made in the nominal pointing position while the observation for ({\em a}) 
was made in one of the offset pointing positions.\label{fig:lc16o}}
\end{figure}

A more careful examination of the 72 light curves with 
high-amplitude variability indicates that there are at
least four types of high-amplitude variability, and 
examples of the most common type are shown in 
Figure~\ref{fig:lc16f} while the other three types are 
shown in Figure~\ref{fig:lc16o}.  The light curves shown 
in Figure~\ref{fig:lc16f} can be interpreted as a series 
of 10--100~s flares; however, in some of the light curves 
the flares occur so often that it is difficult to 
distinguish the individual flares.  We refer to this as 
``flaring'' behavior below, and we chose the examples to 
illustrate that flaring occurs over a wide range of times 
during the outburst as shown in Figures~\ref{fig:lc} and
\ref{fig:parameters}.  While all of the examples shown
in Figure~\ref{fig:lc16f} are from IS observations, flaring
also sometimes occurs during observations classified as
TD and SPL.  Figure~\ref{fig:lc16f} also shows the 
9--20/3--9 keV hardness as a function of time.  For the
observations shown in panels {\em a}, {\em b}, and 
{\em c}, the hardness is tightly correlated with the 
total 3--20~keV rate (i.e., these are hard flares).  
However, this correlation is not seen for the panel
{\em d} observation.

Even more extreme variability with $A_{\rm pp} > 1.0$
is seen in the two light curves shown in 
Figure~\ref{fig:lc16o}a and \ref{fig:lc16o}b.  These 
light curves exhibit deep dips that are clearly different 
from the other types of variability observed.  The dip 
shown in Figure~\ref{fig:lc16o}a is very similar to the 
dip observed from 4U~1630--47 during its 1996 outburst 
\citep{tomsick98,kuulkers98}.  Spectral analysis of the 
1996 dip indicates that it is likely caused by absorption, 
perhaps from accretion disk material, but the fact that 
the dip is not spectrally hard indicates partial covering 
of the X-ray source \citep{tomsick98,kuulkers98}.  
Similarly, the dip shown in Figure~\ref{fig:lc16o}a shows, 
at most, only a moderate level of hardening.

A third type of variability is illustrated in 
Figure~\ref{fig:lc16o}c and can be described as short, 
10--20~s dips, and similar dips have been reported during 
the 1998 outburst from 4U~1630--47 \citep{tk00,dieters00}.  
The hardness ratio indicates that the spectrum becomes 
softer during the dips.  During this outburst, we only 
see these short, soft dips during the three observations
from MJD 52,794.289, 52,795.340, and 52,801.445, and a total 
of 7 or 8 short dips are seen in the 19~ks of exposure 
time accumulated for these three observations, which are
all classified as SPL.

The final type of high-amplitude variability is shown in 
Figure~\ref{fig:lc16o}d and consists of a single, very 
hard flare that occurred at the end of the observation 
performed on MJD 53,087.832, at which time the source 
was in the TD state.  The flare did not occur close to the 
time of any SAA passage, and we checked that the PCA electron 
ratio was not high during the time of the flare, indicating
that the flare did not have any instrumental or 
environmental cause.  In addition, we checked for solar
flares using the X-ray data from the Solar X-ray Imager
(SXI) on the Geostationary Operational Environmental 
Satellites (GOES-12), but the sun did not flare during 
the {\em RXTE} observation.  Finally, we extracted a 
16 s HEXTE light curve, and we see that HEXTE also
detected the hard flare.  Thus, we conclude that origin
of the flare was an astronomical source in the {\em RXTE}
field-of-view.  It is likely that the flare came from
4U~1630--47, but we cannot rule out the possibility that
it came from one of the other sources in the field-of-view, 
such as IGR~J16358--4726 or IGR~J16320--4751.

\section{Examples of Energy and Power Spectra}

\begin{figure}
\centerline{\includegraphics[width=0.52\textwidth]{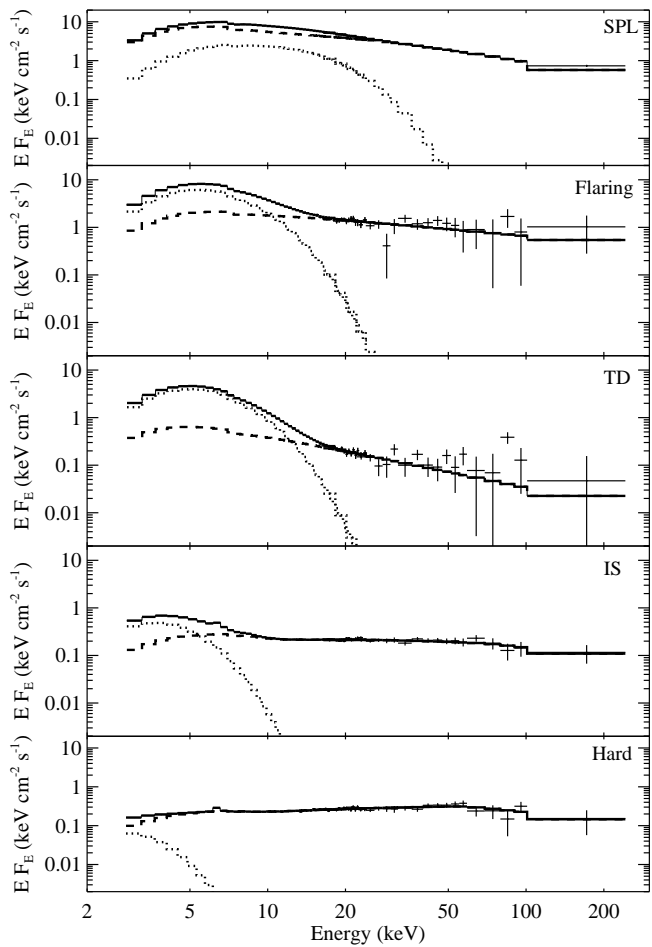}}
\vspace{0.5cm}
\caption{Five example {\em RXTE} energy spectra fitted 
with the model described in the text.  In addition to the total model 
({\em solid line}), the {\em dotted line} shows the disk-blackbody 
component and the {\em dashed line} shows the power-law component.  
From top to bottom, the spectra are ordered by decreasing 3--200 keV 
flux and disk-blackbody temperature, and illustrate examples of the 
following states or behaviors: SPL, Flaring, TD, IS, and Hard.  See 
Table~\ref{tab:spectra} for details such as observation IDs, observation 
times, and parameter values.\label{fig:spectra}}
\end{figure}

\begin{figure}
\centerline{\includegraphics[width=0.52\textwidth]{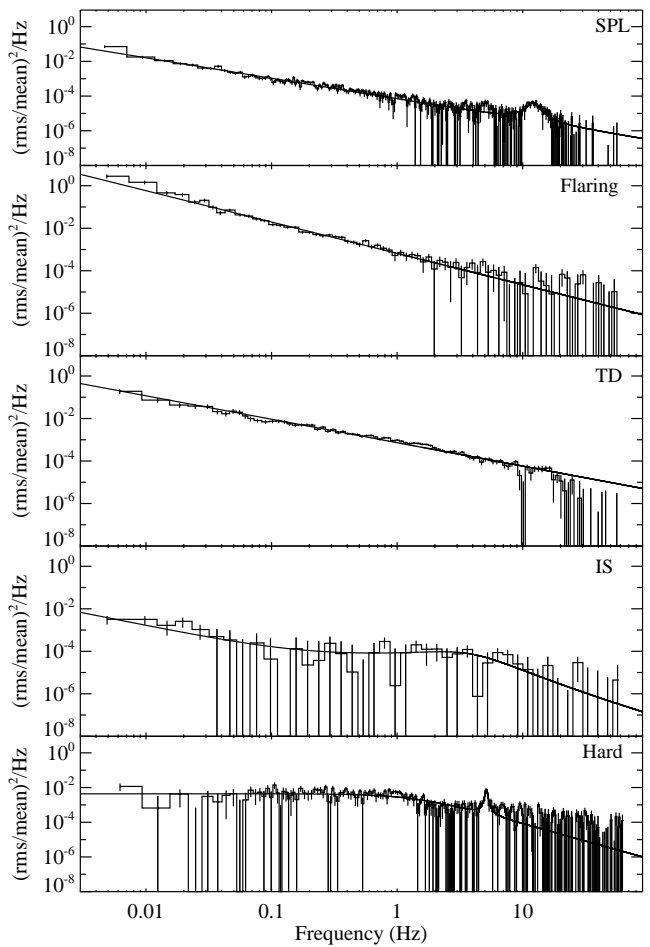}}
\vspace{0.5cm}
\caption{{\em RXTE} power spectra for each of the 
five observations shown in Figure~\ref{fig:spectra}, illustrating 
examples of the following states or behaviors: SPL, Flaring, TD, 
IS, and Hard.  The solid lines represent model fits, and the 
parameters for these models are given in Table~\ref{tab:power}.
\label{fig:power}}
\end{figure}

We selected representative observations for more detailed
spectral and timing analysis.  The selected observations
include one observation from each of the spectral states
(SPL, TD, and Hard), and two observations from the IS
state:  One that exhibits the flaring behavior described
above, and one that does not.  The energy spectra from these 
observations are shown in Figure~\ref{fig:spectra}, and the 
parameters from the spectral fits are given in Table~\ref{tab:spectra}.  
From the SPL state to the Hard state, the spectra are ordered 
according to decreasing $kT_{\rm in}$ and 3--200 keV flux, 
as shown in Table~\ref{tab:spectra}.  While $kT_{\rm in}$ is 
correlated with the flux, the disk-blackbody normalization 
is anti-correlated with the flux.  The changes for $\Gamma$
and $PLR$ do not have a simple relationship with the total 
flux.  The $PLR$ is the highest at the highest flux and the 
lowest flux, but the thermal, disk-blackbody, component 
contributes a higher fraction of the flux at intermediate
flux levels.  Table~\ref{tab:spectra} also provides the 
measured values for the column density ($N_{\rm H}$).  The 
values, with 90\% confidence error bars, indicate that the 
column density drops significantly with flux.

The 0.005--64~Hz rms-normalized power spectra for the same 
five observations are shown in Figure~\ref{fig:power}.  In 
fitting the power spectra, we considered various combinations 
of three different components, which we also previously used 
to fit the 4U~1630--47 spectra from the 1998 outburst 
\citep{tk00}: A power-law function (PL); a band-limited noise 
component (BL), which we modeled as a zero-centered Lorentzian; 
and a QPO, which we modeled using a Lorentzian.  For the SPL 
state, the 0.005--10~Hz continuum is well-described by the PL 
model.  In addition, a QPO is present at $12.15\pm 0.15$ Hz 
with a quality factor of $Q = 4.4\pm 0.7$ and a fractional rms 
of 1.30\%$\pm$0.07\% (see Table~\ref{tab:power}).  Above this 
QPO, the power drops much more rapidly than the extrapolation 
of the PL.  For the IS observation with flaring, nearly 
the entire 0.005--64~Hz range is well-described by the PL.  
There are no QPOs in this case, but there is excess power 
below $\sim$0.02~Hz due to the flaring.  The TD state power 
spectrum has approximately a power-law shape from 0.005--10~Hz.  
There appears to be a narrow dip in the power at 10~Hz, and the 
power drops off more rapidly than the extrapolation of the 
power-law above 20~Hz.  In the IS state, there is evidence for 
BL and PL components, although the statistics are rather poor 
for this state due to a lower count rate and a lower rms noise 
level.  There is no evidence for the presence of QPOs in the 
IS state.  Finally, in the Hard state, the continuum is 
well-described by only the BL component, and a QPO is present 
at $5.13\pm 0.01$ Hz with $Q = 21\pm 5$ and a fractional rms
level of 5.4\%$\pm$0.3\%.  The other Hard state observation 
exhibited a very similar power spectrum but with the QPO at 
$4.03\pm 0.01$ Hz.

The power spectra for the observations we previously identified
as being in the TD, SPL, and Hard states based on their 
spectral properties are consistent with the typical power 
spectra expected for these states as described by \cite{mr03}.  
TD power spectra typically have 0.1--10 Hz rms values $\lsim$6\%
with weak or no QPOs and a power-law shape, and these properties
are consistent with our example TD power spectrum.  For SPL power 
spectra, the continuum often (but not always) has a power-law 
shape, and low-frequency (1--15 Hz) QPOs are usually present.  
Our example SPL power spectrum exhibits similar properties.
Based on the fact that the QPO in our SPL example is relatively
broad, it would be classified as type A or B (depending on its 
phase lag properties) using the QPO classifications given in 
\cite{remillard02}, and the combination of a power-law
continuum and type A or B QPOs is typical \citep{casella04}.

The power spectra for the two observations identified as 
being in the Hard state have relatively strong band-limited
noise, which is a characteristic of Hard state power spectra.
The 0.1--10 Hz rms values for these two observations are 
9.6\%$\pm$0.4\% and 9.0\%$\pm$0.6\%, lying just below the 
10--30\% range that is typical for the Hard state.  While
this could indicate that the source was close to but not
quite in the Hard state, it is important to note that these
are the only two observations for which we found strong 
band-limited noise.  We produced power spectra for the five 
IS observations closest in time to the two Hard state 
observations.  We fitted them with a power-law model and
obtained reduced-$\chi^{2}$ values between 0.4 and 2.1
for 17 dof.  The power-law fits give 0.1--10 Hz rms values
between 1\% and 4\%, indicating that the noise level is
much lower for these observations than for the two 
observations identified as Hard state observations.  Thus, 
these two observations distinguish themselves from the
other observations based on their spectral and timing
properties, which appears to indicate that the source
did briefly enter the Hard state.

The properties of the QPOs seen for the two Hard state
power spectra are also different from the QPO seen in
the SPL.  In addition to being at lower frequency 
4--5 Hz, the Hard state QPOs are much narrower, and
they would be classified as type C QPOs.  The combination
of type C QPOs and band-limited noise is typical
\citep{casella04}.  It is also notable that the 
Hard state power spectra are very similar to those
seen at the end of the 1998 outburst from 4U~1630--47
\citep{tk00}.  As the source declined in 1998, the 
QPO was first detected at 3.4 Hz and then gradually
dropped to 0.2 Hz.  Thus, the 4--5 Hz QPO we see in 
the 2002--2004 outburst likely indicates that a similar 
phenomenon began to occur but was stopped, perhaps by 
an increase in the mass accretion rate.

\section{Discussion}

Good X-ray coverage of 4U~1630--47 during the 2002--2004 
outburst has allowed us to study its source properties 
throughout the outburst in detail, and here, we discuss 
these properties in the context of previous outbursts from 
4U~1630--47 as well as in the context of accreting black 
holes in general.  The 2002--2004 outburst is, by far, the 
longest and brightest that has occurred during the {\em RXTE} 
era, and, in \S$8.1$, we make comparisons to historical 
outbursts.  In our study, we have found several extreme and 
unusual source properties, and, in \S$8.2$, we discuss our
measurements of the soft component, including the extremely 
high inner disk temperatures that occur for some of the Steep 
Power-Law state observations.  In \S$8.3$, we extend the 
analysis described above to constrain the nature of the 
high-amplitude flaring.  In \S$8.4$, we compare the X-ray
and radio properties of the 2002--2004 outburst to those of
the 1998 outburst and discuss radio/X-ray connections.
Finally, our study has provided a test of the quantitative
\cite{mr03} state definitions, and, in \S$8.5$, we discuss
some of the pros and cons of these definitions.

\subsection{Comparison of the 2002--2004 Outburst to Previous Outbursts}

Prior to the 2002--2004 outburst from 4U~1630--47, four outbursts 
from this source had been observed by the {\em RXTE}/ASM during the 
{\em RXTE} era (1996--present).  The mean duration of these four 
outbursts as measured by the {\em RXTE}/ASM is 140 days, and the 
mean peak ASM count rate is 28 c/s (0.38 Crab).  From the ASM light 
curve (Figure~\ref{fig:lc}a), the duration of the 2002--2004 
outburst is 825 days, and the peak ASM rate is 62 c/s (0.84 Crab), 
making it longer by a factor of nearly six and brighter by a factor 
of 2.2 than the mean values for the first four outbursts.  

While the current outburst is unusual compared to the 4U~1630--47 
outbursts of the past decade, it is not unprecedented in duration 
or in brightness when compared to the entire group of outbursts 
going back to 1969.  Although there are many cases where poor X-ray 
coverage makes it difficult to tell whether outbursts were extended
or not, the clearest example of an extended outburst occurred when 
the {\em Ginga} All-Sky Monitor detected the source for 2.4 years 
between 1988 October and 1991 March \citep{kuulkers97}.  The 
{\em Ginga} light curve shows that the 1988--1991 outburst had many 
similarities to the current outburst, including flares where the 
flux reached $\sim$0.6 Crab as well as multiple time periods of 
low flux during the outburst.  For the 1988--1991 outburst, the 
low flux periods are separated by $\sim$220 days \citep{kuulkers97}.  
The ASM light curve for the 2002--2004 outburst has local minima at 
MJD 52,685, 53,000, 53,075, and 53,250, indicating separations of 315, 
75, and 175 days; thus, they are the same order of magnitude as the 
220 day separations, but they are clearly different.

With a peak flux of 0.84 Crab (1.5--12 keV), the 2002--2004
outburst is somewhat brighter than the 0.6 Crab (1--20 keV) 
flares detected during the 1988--1991 outburst.  However, at 
a 3--6 keV flux of 1.4 Crab \citep{csl97}, the 1977 outburst, 
which lasted for about 0.3 years, was brighter than the 
current outburst.  In summary, while the 2002--2004 outburst 
is one of the longest and brightest outbursts ever detected 
from 4U~1630--47, it is not unprecedented in either category.  
On the other hand, no previous outburst was both brighter 
and longer than the current outburst, so it is very likely
that the total mass accreted is higher for the 2002--2004
outburst than for any previous outburst.

The high level of recent activity from 4U~1630--47 strengthens 
the argument made by \cite{csl97} that the mass accretion rate 
from the binary companion ($\dot{M}_{\rm c}$) is unusually high 
for this source.  Further, \cite{csl97} point out that this 
implies a very long binary orbital period ($P_{\rm orb} > 12$ 
days) for 4U~1630--47 based on the calculations of 
\cite{vanparadijs96}, which show that, for a given $P_{\rm orb}$, 
an X-ray binary will only be transient if $\dot{M}_{\rm c}$ is 
smaller than a critical value.

\subsection{The Soft Component: Disk Temperatures and Luminosities}

In BHC energy spectra, the presence of a strong soft component
is a clear indication that we are seeing thermal emission from
an optically thick accretion disk.  The basic physical properties
that determine the shape of the soft component include the mass
accretion rate, the mass of the black hole, the inner radius of
the disk, and the binary inclination.  If we could assume a
standard \cite{ss73} accretion disk, at least some of these
parameters might be directly measurable by modeling the shape
of the soft component; however, in practice, other physical
processes can be important and can complicate the interpretation
of any derived parameters.

\begin{figure}
\centerline{\includegraphics[width=0.52\textwidth]{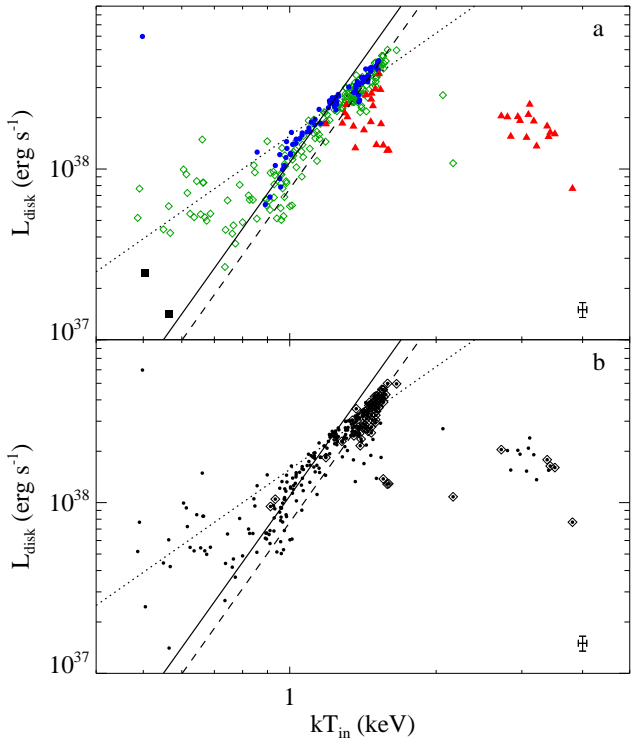}}
\caption{For each {\em RXTE} observation, the figures
(panels {\em a}) and ({\em b}) show $L_{\rm disk}$, the bolometric 
disk luminosity derived from the diskbb parameters assuming a distance 
of 10~kpc and a binary inclination of $60^{\circ}$, vs. the inner disk 
temperature, $kT_{\rm in}$.  The colors/symbols in ({\em a}) represent 
the different spectral states (Red triangles = SPL, Green open 
diamonds = IS, Blue circles = TD, Black squares = Hard).  
Representative errors are shown in the bottom right hand corner of 
the plots.  In ({\em b}), the observations with high-amplitude 
variability are marked with diamonds.  In ({\em a}) and ({\em b}),
the {\em solid and dashed lines} are lines of constant $R_{\rm in}$ 
($L_{\rm disk}\propto T_{\rm in}^{4}$), while the {\em dotted line} 
indicates a $L_{\rm disk}\propto T_{\rm in}^{2}$ relationship (see 
text for a discussion of the physical significance of these 
relationships).\label{fig:dl}}
\end{figure}

For 4U~1630--47, we detect the soft component over a wide range
of luminosities and mass accretion rates.  While we have modeled
the soft component using the disk-blackbody (diskbb) model, the 
limitations of this model must be understood when interpreting 
the parameters.  For example, the shape of the soft component
can be drastically changed if the disk opacity is dominated
by electron scattering rather than free-free absorption
\citep{st95}.  It has been shown that this effect can cause
measurements of inner disk radii to be underestimated by a 
factor of five or more \citep{mfr00}.  In addition, at high mass 
accretion rates, additional cooling mechanisms may cause a 
change from the thin \cite{ss73} disk solution to a geometrically 
thicker ``slim'' disk \citep{abramowicz88}.  This can lead to
significantly more material, and thus emission, at small radii
and can also produce a much flatter radial temperature profile
($T\propto R^{-p}$), with a change in $p$ from 0.75 to $\sim$0.5
\citep{watarai00}.  As a final example of the limitations of
the diskbb model, it has been shown that assumptions about
the boundary conditions at the inner radius of the disk can
be important.  The nonzero torque boundary condition assumed
(basically for computational convenience) in the diskbb model
can lead to an overestimation of the inner disk radius by a
factor of more than two \citep{zimmerman05}.

Using observations of BHC systems XTE~J1550--564, GRO~J1655--40, 
LMC~X-3, and observations from previous outbursts of 4U~1630--47, 
Kubota and co-workers show how effects of electron scattering and
changes in the radial temperature profile can manifest themselves 
\citep{km04,kme01,afk04}.  Following these studies, we plot in
Figure~\ref{fig:dl} the disk temperature ($kT_{\rm in}$) vs. the 
bolometric disk luminosity ($L_{\rm disk}$) as derived from the 
4U~1630--47 diskbb parameters.  In deriving the luminosity, we 
assume a source distance of 10~kpc and a binary inclination
of $60^{\circ}$, but these are highly uncertain.  In 
Figure~\ref{fig:dl}, we plot a solid line representing the slope 
of the $L_{\rm disk}\propto T_{\rm in}^{4}$ relationship that is 
expected for a standard \cite{ss73} accretion disk with a constant 
inner radius ($R_{\rm in}$).  A large number of mostly TD and IS 
points at disk luminosities between $3\times 10^{37}$ and 
$3\times 10^{38}$ erg~s$^{-1}$ lie close to the line of constant 
$R_{\rm in}$, and, for these observations, we may be seeing a 
standard disk with a relatively stable inner radius.  However, 
many of the points also deviate from this line, and we identify 
three regions of deviation that likely have distinct explanations.  
First, several IS and Hard state observations at the lower 
luminosities ($\lsim$$10^{38}$ erg~s$^{-1}$) show disk temperatures 
well below values that would be consistent with the line of 
constant $R_{\rm in}$.  These are cases for which the overall 
source luminosity and presumably also the mass accretion rate are 
low, and these are likely cases where the inner disk radius 
increases or at least where the inner part of the disk is 
radiatively inefficient.  

A second region of deviation includes TD, IS, and some of the
SPL observations and occurs at the highest disk luminosities.
This flattening of the $L_{\rm disk}$-$kT_{\rm in}$ relationship
may be similar to what has been seen for XTE~J1550--564 at 
high $L_{\rm disk}$ \citep{km04}.  For XTE~J1550--564, \cite{km04} 
showed that the relationship flattened to a slope close to
$L_{\rm disk}\propto T_{\rm in}^{2}$, which could be an indication
of a transition to a slim disk based on the work of \cite{watarai00}.  
In 4U~1630--47, it is clear from Figure~\ref{fig:dl} that the 
source leaves the solid line above $\sim$$2\times 10^{38}$ 
erg~s$^{-1}$, and it appears that it may begin to follow the
$T_{\rm in}^{2}$ relationship (the dotted line in Figure~\ref{fig:dl})
for at least some luminosity range.  However, the source appears
to deviate from the $T_{\rm in}^{2}$ relationship at the very
highest values of $L_{\rm disk}$, and it is possible that the
source recovers the $T_{\rm in}^{4}$ relationship (the dashed 
line in Figure~\ref{fig:dl}).  Although the exact evolution
is not completely clear, it is interesting that the observations
at the highest values of $L_{\rm disk}$ that deviate from the
solid line are also the observations for which the high-amplitude
flaring occurred (see Figure~\ref{fig:dl}b).  For transitions 
between the standard disk and the slim disk, theory predicts a 
limit cycle with a region of instability in between the two 
solutions.  Thus, the flaring may be a consequence of the limit 
cycle, and below this is a possibility we explore further.

A third region of deviation from the line of constant $R_{\rm in}$
is contains mostly SPL observations for which $kT_{\rm in}$ is
extremely high, including 15 observations with $kT_{\rm in}$
between 2.7 and 3.8~keV.  Along with extremely high temperatures, 
the spectra exhibit very low values of $N_{\rm diskbb}$ in the 
range of 1.2--12.3, implying values of $R_{\rm in}$ that are
unphysical.  As an example, for a source distance of 10~kpc and
a binary inclination of $60^{\circ}$ (as assumed above), this
range of normalizations indicate inner radii between 1.5 
and 5.0 km; the former being an order of magnitude lower than
the gravitational radius of a 10\Msun~black hole.  Rather than
extremely small inner disk radii, these high temperatures and
luminosities are much more likely to be caused by spectral 
hardening due to the dominance of electron scattering in
the inner region of the accretion disk.  

Although the explanation for the high SPL state values of 
$kT_{\rm in}$ is very likely electron scattering, it is notable
that the extremely high temperatures are seen for such a large
number of observations.  During its 1998 outburst, 4U~1630--47 
entered the SPL state, and fits to {\em RXTE} spectra gave 
$kT_{\rm in} = 1.6$--1.7 keV and $N_{\rm diskbb} = 46$ 
\citep{tk00,mr03}, which are considerably less extreme when 
compared to the 2002--2004 values.  Very high values of 
$kT_{\rm in}$ have been seen for other accreting BHCs, although 
they are not common.  For the ten SPL spectra of accreting 
black holes studied by \cite{mr03}, only XTE~J1550--564 and 
GRO~J1655--40 have $kT_{\rm in} > 2.0$ keV.  In 1998, the 
XTE~J1550--564 spectrum showed $kT_{\rm in} = 3.3$ keV and 
$N_{\rm diskbb} = 7.8$ \citep{sobczak00,mr03}, which are within 
the range of values we see for 4U~1630--47; however, for 
XTE~J1550--564, this spectral shape was only seen for a single 
{\em RXTE} observation that occurred during a remarkable 6 Crab 
flare, during which powerful superluminal jets were ejected 
\citep{hannikainen01}.  For GRO~J1655--40, another superluminal 
jet source, \cite{mr03} give a SPL example where the disk-blackbody 
temperature is 2.2~keV, and this source showed temperatures of 
$\sim$2 keV for a few other observations \citep{sobczak99}.  
However, for the other eight \cite{mr03} SPL systems, 
$kT_{\rm in}$ is in the range 0.5--1.7 keV.  Although not 
discussed in \cite{mr03}, very high disk-blackbody temperatures 
have also been seen for GRS~1915+105.  For six of the 1996--1997 
observations made when the GRS~1915+105 luminosity was very 
high, \cite{mmr99} report values of $kT_{\rm in}$ in the range 
2.6--4.8~keV.  Based on the \cite{mr03} definitions and the 
parameters reported in \cite{mmr99}, GRS~1915+105 was in the 
SPL state during these observations, but it should be noted
that the properties of GRS~1915+105 make it difficult to
classify its behaviors into the canonical spectral states
\citep{reig03}.

Concerning 4U~1630--47, we can conclude that the extremely 
high disk-blackbody temperatures that we measure during the 
2002--2004 outburst are rare and may be a new phenomenon for 
this source.  When compared to other BHC sources, the 
4U~1630--47 temperatures are only matched by XTE~J1550--564 
and GRS~1915+105.  For 4U~1630--47 and XTE~J1550--564, the 
high temperatures are only measured during observations for 
which the sources are at their very brightest, and, for 
GRS~1915+105, the high temperatures occur when the source 
is at or close to its brightest level \citep{mmr99,dwg04}.  
It is possible that these high temperatures are an indication 
of the highest accretion rates that are possible from these 
systems.  As described above, the cause of these high 
temperatures may be electron scattering in the inner disk.
This causes the observed temperature, $kT_{\rm in}$, to be 
higher than the effective temperature, $kT_{\rm eff}$, by a 
factor of $f$.  For the highest value of $kT_{\rm in}$ that 
occur for 4U~1630--47 and GRS~1915+105 ($\sim$4 keV), $f$ 
would need to be $\sim$3 to obtain the temperatures expected 
for a \cite{ss73} disk around a 10\Msun~black hole.  Although 
this is higher than the value of $f = 1.7$ theoretically
expected for luminosities around 10\% of the Eddington 
luminosity \citep{st95}, these authors also find that much 
higher values of $f$ can occur for luminosities approaching 
the Eddington limit.  

\begin{figure}
\centerline{\includegraphics[width=0.52\textwidth]{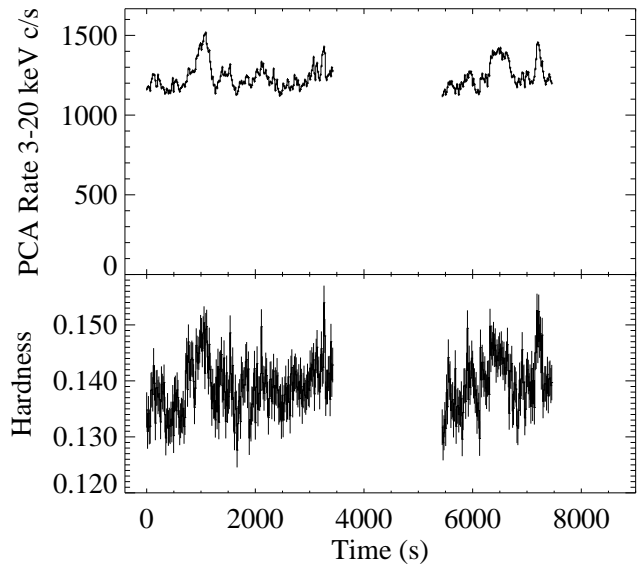}}
\vspace{-0.3cm}
\caption{The PCA light curve and 9--20/3--9 keV 
hardnesses vs. time with 16~s time bins, showing that flaring
was also seen during one of the previous 4U~1630--47 outbursts.
These data come from an {\em RXTE} observation made on 2000
November 18 (MJD 51,866.9) during the 2000--2001 outburst 
(Observation ID 50120-01-04-00).\label{fig:lc_50120}}
\end{figure}

\subsection{Flaring Behavior}

Flaring from accreting BHC systems is seen on a wide range 
of time scales and may have various physical origins, 
including (but not limited to) accretion of clumps of matter
or magnetically powered particle acceleration.  While black 
hole variability is common, the flaring we see for 4U~1630--47, 
with amplitudes as high as $A_{\rm pp} = 1.0$ on
time scales of 10--100~s, is extreme.  The 4U~1630--47 
amplitudes are comparable to the wild variability seen for 
GRS~1915+105 \citep[e.g.,][]{belloni00}; however, the 
4U~1630--47 light curves do not show the distinctive repeating 
patterns seen for GRS~1915+105.  Recently, high amplitude 
variability at relatively long time scales has been seen for 
the black hole systems XTE~J1859+226 \citep{casella04} and 
H~1743--322 \citep{miller04,homan05}.  Although the flares in 
these systems are not as extreme as we see in 4U~1630--47, 
they may be related.

To determine if the flaring is unique to the 2002--2004 
outburst from 4U~1630--47, we inspected the 16~s light curves
for the nearly 300 pointed {\em RXTE} observations made during 
the four previous outbursts.  Similar flaring only occurred
for three of the observations, and these observations were
made during the 2000--2001 outburst over the time period
2000 November 16--18.  Figure~\ref{fig:lc_50120} shows the 
16~s light curve and hardness ratio vs. time for the
November 18 observation, and it is notable that the PCA
count rates and hardness levels are similar to those seen
during the 2002--2004 flaring observations.  For the 
November 18 observation, we extracted a PCA plus HEXTE
energy spectrum and fitted the spectrum as described above
for the 2002--2004 observations.  The spectral parameters
are remarkably similar to those seen for the 2002--2004
flaring observations.  The measured inner disk temperature
is $kT_{\rm in} = 1.450^{+0.007}_{-0.010}$ keV, and 
$L_{\rm disk} = 3.2\times 10^{38}$ erg~s$^{-1}$, putting it
in the same region as the other flaring observations in
Figure~\ref{fig:dl}.  Also, $\Gamma = 2.59^{+0.04}_{-0.02}$
and $PLR = 0.43$, so the observation would be classified
as IS.

\begin{figure}
\centerline{\includegraphics[width=0.52\textwidth]{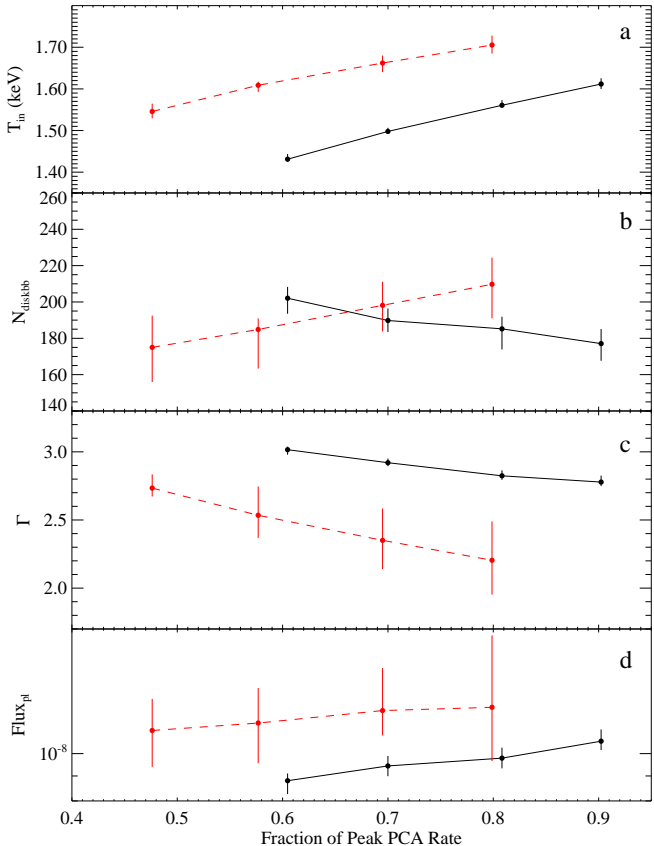}}
\vspace{0.3cm}
\caption{The spectral parameters measured for two of the 
flaring observations as a function of the PCA count rate divided by the
peak PCA count rate.  The red points connected with dashed lines and
the black points connected with solid lines are for observation IDs
70113-01-43-00 and 80117-01-13-02, respectively, and the light curves
for these observations are shown in Figure~\ref{fig:lc16f}.  ({\em a}) 
and ({\em b}) show the disk-blackbody parameters, while ({\em c}) and
({\em d}) show the power-law parameters, including the flux of the
power-law component in units of erg~cm$^{-2}$~s$^{-1}$.  For both 
observations, the two clear trends are an increase in $kT_{\rm in}$
and a hardening of $\Gamma$ with rate.\label{fig:flaring}}
\end{figure}

For the observations with light curves shown in 
Figures~\ref{fig:lc16f}b (70113-01-43-00) and \ref{fig:lc16f}c 
(80117-01-13-02), we performed spectral fits to study the 
spectral evolution as a function of PCA count rate.  In both 
cases, we divided the 16~s time resolution data into different 
PCA count rate ranges.  We separated the full range from minimum 
rate to maximum rate into four sub-ranges of equal size and 
produced four PCA spectra.  We did not use HEXTE for this 
analysis because the HEXTE rocking would complicate the analysis.  
We fitted the four spectra simultaneously, leaving the parameters 
for the iron features free, but requiring that they be the same 
for all four spectra.  We did not include a high energy 
cutoff as a cutoff was not required for either observation.
Originally, for both observations, we allowed the column density 
to be different for the four spectra.  For 70113-01-43-00, when
we forced $N_{\rm H}$ to be the same for all four spectra, the
quality of the fit changed from $\chi^{2}/\nu = 187.8/174$ to
$\chi^{2}/\nu = 189.4/177$.  For 80117-01-13-02, the change was 
from $\chi^{2}/\nu = 208.8/174$ to $\chi^{2}/\nu = 209.7/177$, 
indicating that the spectra are consistent with a constant 
$N_{\rm H}$ for both observations.  Figure~\ref{fig:flaring}
shows the evolution of the spectral parameters with count
rate.  For both observations, clear trends are seen with
$kT_{\rm in}$ increasing and $\Gamma$ hardening with count
rate.  The results indicate that both the soft and hard 
components are affected.  In light of the above standard/slim
disk discussion above, perhaps the most important result is
the clear and strong increase in $kT_{\rm in}$.  The 
temperature increase is expected if the disk solution 
changes from a standard disk at low count rates to a slim
disk at high count rates.  This, along with the fact that
the flaring may be a consequence of the zone of instability
between the standard and slim disk solutions, make this 
explanation attractive.  However, from the spectral evidence 
alone, we cannot rule out that the disk temperature rises 
because of an increase in the mass accretion rate.

Although the spectral analysis indicates that changes in 
the accretion disk are important in producing the flaring, 
one might also ask whether the flaring behavior could 
have any physical connection to outflows or jets in the
system.  For example, in the case of the BHC H~1743--322, 
from which spectrally hard flares were also recently 
detected, the flaring was accompanied by the presence of 
blue-shifted absorption lines detected in spectra taken 
with the {\em Chandra X-ray Observatory} \citep{miller04}.  
As these authors point out, the blue-shifting of the lines 
strongly suggest the presence of an outflow.  In addition, 
\cite{miller04} find that the line strength varies during 
the $\sim$300~s variations seen in H~1743--322, and 
\cite{homan05} suggest the possibility that, if the outflow 
is energetic enough, it may be able to produce hard emission 
via up-scattering.  For 4U~1630--47, we do not have information 
about the presence of absorption lines, but, as our spectral 
analysis indicates that the flaring is connected to the 
accretion disk, a link between the flaring and an outflow 
would represent an important disk/jet connection.

One might expect that an outflow energetic enough to lead to 
strong flares would also lead to a high level of radio emission.  
However, for H~1743--322, the radio flux is actually lower 
during {\em Chandra} observations when the absorption lines
are detected than when they are not detected \citep{miller04}.  
For 4U~1630--47, \cite{hannikainen02} observed the source in the 
radio band on three occasions (2002 September 16, 19, and 24) 
when the source was exhibiting flaring, and they obtained upper 
limits on the 3~cm and 6~cm radio flux of 95, 180, and 
180~$\mu$Jy, respectively, for the three dates.  This does not 
preclude the presence of an outflow, but it does place limits 
on how strong any outflow might be.  We note that several other 
radio observations of 4U~1630--47 occurred during various parts 
of the 2002--2004 outburst (including times of other flaring 
episodes and when the source was in the SPL state), but no radio 
detections were reported.

\subsection{A Connection between X-Ray Properties and Radio 
Jet Emission}

\begin{figure}
\centerline{\includegraphics[width=0.52\textwidth]{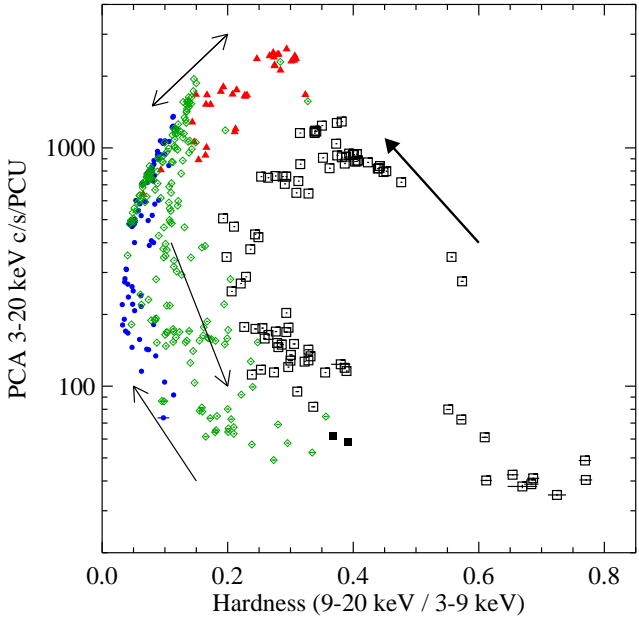}}
\caption{Hardness-intensity diagram showing the 
3--20 keV PCA count rates vs. the 9--20/3--9 keV hardness ratio
for the {\em RXTE} observations from the 2002--2004 and 1998 
outbursts.  The 1998 outburst is the only time when radio 
emission was detected from 4U~1630--47.  The 2002--2004 
observations are marked using the same colors and symbols
as described in the captions of Figures~\ref{fig:spectra} and 
\ref{fig:dl}.  The 1998 observations are marked with open black 
squares.  The thick-lined arrow with the filled arrowhead shows 
the counter-clockwise motion seen for the 1998 outburst.  For 
the 2002--2004 outburst, the pattern of the motion is complicated 
as shown by the thin-lined arrows with the un-filled arrowheads.
\label{fig:hi}}
\end{figure}

The absence of radio emission during the 2002--2004 outburst
is also interesting when comparing the X-ray properties during
this outburst to those seen during the 1998 outburst, which is 
the only time radio emission has been detected from 4U~1630--47.  
In Figure~\ref{fig:hi}, for both the 2002--2004 and 1998
outbursts, we show the hardness, defined as the ratio of the 
9--20~keV PCA count rate to the 3--9~keV PCA count rate, vs. 
the source intensity (the 3--20~keV PCA count rate).  For the 
2002--2004 outburst, we divide the observations 
into the different spectral states.  All of the 1998 observations 
are marked with open squares.  There are potentially important 
differences between the X-ray properties for the two outbursts 
in light of the fact that radio jet emission was only present for 
the 1998 outburst.  Although both outbursts have ``c''-shaped 
hardness-intensity diagrams, as is relatively typical for BHC 
systems \citep{fbg04,hb04}, the 1998 outburst is shifted in 
hardness, indicating that the spectrum was harder in 1998 than 
in 2002--2004.  Also, as marked in Figure~\ref{fig:hi}, 
4U~1630--47 traveled through the hardness-intensity diagram in 
a counter-clockwise fashion during its 1998 outburst, while, in 
2002--2004, the source often moved in the clockwise direction.  
A related fact is that 4U~1630--47 entered a hard and bright 
state at the beginning of its 1998 outburst, whereas in 2002--2004, 
there is no evidence that this occurred.  Although it is likely 
that 4U~1630--47 did enter a hard state during the 2002--2004 
rise as this is the common pattern in BHC systems, the 
combination of the ASM and pointed {\em RXTE} observations (see 
Figure~\ref{fig:lc}) indicate that such a state would have had 
to last a very short time and be limited to a time period when 
the source was dim.

These patterns are especially interesting in light of recent
work on connections between black hole X-ray states, a source's 
position in the hardness-intensity diagram, and radio jet 
ejections \citep{hb04,corbel04,fbg04,fb04}.  A general pattern 
seen in a number of BHC outbursts is that the systems will evolve 
from the Hard state to an intermediate state and then a radio 
ejection will occur during the subsequent transition to the SPL 
state \citep{corbel04}.  \cite{fbg04} use the hardness-intensity 
diagram to quantify this effect, and it should be noted that they 
follow \cite{hb04} by using somewhat different terminology, 
describing the transition from the IS to the SPL states as a 
transition from the ``Hard Intermediate'' state to the ``Soft 
Intermediate'' state.  \cite{fbg04} suggest that each BHC system 
has a threshold hardness and that major jet ejections are only 
produced when the source crosses this ``jet-line'' from a high 
to a low hardness level.  If this is the case, Figure~\ref{fig:hi} 
suggests that the jet-line for 4U~1630--47 may be around a hardness 
of 0.4.  Our results for 4U~1630--47 provide evidence in favor 
of the jet-line concept; however, based on the behavior of 
4U~1630--47, it is not entirely clear if the presence of a 
radio jet in 1998 occurred because the outburst was harder 
overall or if the source simply entering a bright Hard state 
led to the jet ejection.

\subsection{Notes on Spectral States}

Our analysis of this data also provides a test of the 
quantitative spectral state definitions
of \cite{mr03}.  In some ways, these definitions are quite
successful.  For example, it is impressive that the only 
two observations with spectral parameters meeting the
Hard state requirements are also the only two we found
with power spectra that include a strong band-limited
noise component.  Also, looking at the hardness-intensity 
diagram in Figure~\ref{fig:hi}, these two observations lie 
at the extreme end of the c-shaped pattern.  However, in 
other areas, the definitions appear to be less satisfactory.
Although most of the observations labeled as SPL lie
at the other end of the c-shaped pattern in the 
hardness-intensity diagram, the SPL observations have
a extremely wide variety of properties.  For example,
for about half of the SPL observations $kT_{\rm in}$
lies in the 2.7--3.8~keV range, while the other half
have temperatures $<$1.8~keV.  We argue above that
there are important physical differences between these
two groups.  Also, it is notable that over half of the
observations made during the 2002--2004 outburst are
put in the IS, meaning that a large fraction of the
observations have X-ray properties that do not meet
the requirements for any of the \cite{mr03} states.
Overall, while the \cite{mr03} criteria appear to be
useful for classifying the observations at the extremes 
of BHC behavior, it should be recognized that they, by 
themselves, do not provide a complete description of
a BHC outburst due to significant variations in 
properties within states and the large fraction of IS 
observations.

\section{Summary and Conclusions}

Outstanding {\em RXTE} coverage of 4U~1630--47 during
its 2002--2004 outburst has allowed us to study the
detailed evolution of its X-ray spectral and timing 
properties over a period of more than 2 years.  
Historically, this outburst is among the longest and
brightest seen in 36 years of observing 4U~1630--47, 
and it is very likely that it is the largest ever
observed in terms of total mass transfer.

The X-ray properties during this outburst were also
extreme, including 15 observations with very high 
disk-blackbody inner disk temperatures between 2.7
and 3.8~keV.  The inner disk radii inferred from 
these fits are unphysically small, and it is likely 
that the high temperatures and small radii are 
caused by electron scattering.  This explanation
requires a spectral hardening factor of $f\sim 3$, 
implying a source luminosity that is considerably 
higher than 20\% of the Eddington limit \citep{st95}, 
which is not unreasonable as we measure 3--200~keV
luminosities of $5\times 10^{38}$ erg~s$^{-1}$
($d$/10 kpc)$^{2}$.

At the highest disk luminosities, we detect a deviation
from the $L_{\rm disk}\propto T_{\rm in}^{4}$ relationship
(line of constant $R_{\rm in}$) seen at lower luminosities
as well as high-amplitude flaring.  The deviation may be a 
sign of a transition from a standard disk to a slim disk as 
suggested by Kubota and co-workers.  The flaring behavior 
of 4U~1630--47 may be consistent with this interpretation
as a zone of instability is expected between the two
disk solutions.  Also, our spectral analysis of flaring
observations, showing that $kT_{\rm in}$ is correlated
with PCA count rate is consistent with a change from
a standard disk to a slim disk.

Although sensitive radio observations occurred during
the 2002--2004 outburst, no strong radio emission that
would indicate the presence of radio jets was detected.
This is interesting in light of the fact that the X-ray 
properties were very different during the 1998 outburst 
when radio jet emission was detected.  Compared to the 
2002--2004 outburst, the 1998 outburst hardness-intensity 
diagram was shifted to a higher hardness level, and, in 
1998, the source entered into a bright and hard state, 
while it did not in 2002--2004.  These findings support 
the connections between radio jets and spectral states 
found by \cite{corbel04} and the jet-line idea recently 
proposed by \cite{fbg04}.  

Finally, our analysis of a large number of {\em RXTE}
observations has provided a good test of the quantitative
\cite{mr03} spectral state definitions.  While the Hard 
state appears to be well-defined, the spectral and timing 
properties of the observations selected as SPL are highly
non-uniform.  Also, it is notable that over half of the 
observations are put in the IS because they do not meet 
the requirements of any of the \cite{mr03} definitions.
The results show that 4U~1630--47 exhibits many properties 
not encompassed by the \cite{mr03} definitions that are 
likely to be physically important.

\acknowledgments

JAT would like to thank Tomaso Belloni, Ken Ebisawa, 
Joern Wilms, Rick Rothschild, and Katja Pottschmidt for 
useful discussions.  JAT thanks Jerome Rodriguez, 
Luigi Foschini, and Katja Pottschmidt for help with 
the {\em INTEGRAL} data analysis.  JAT acknowledges 
partial support from NASA grants NAG5-12703 and NNG04GA49G.
We thank the referee, Tomaso Belloni, for a useful report
that helped us to improve the manuscript. 

\begin{center}
{\bf\normalsize Appendix A}
\end{center}

As described in \S$5$ above, there are four {\em RXTE} observations
for which we see strong positive residuals at high energies 
($\gsim$30 keV) after fitting the PCA plus HEXTE spectra with our 
standard spectral model.  These occur for observations made at the 
nominal pointing position, and during these observations, 
IGR J16320--4751 was in the field-of-view (FOV), $0^{\circ}.58$ 
from 4U~1630--47 and the center of the FOV.  Due to the high level
of activity from IGR J16320--4751, its known hard spectrum, and
its known strong variability 
\citep[see][and references therein]{foschini04}, we suspect that 
this source may be producing much of the high energy emission that
we see in the four observations with strong positive residuals.
As a test, we compare the flux of the high energy emission seen
in one of these observations (Observation ID 70417-01-07-02 taken 
on MJD 52,539.094) to measurements of IGR J16320--4751 taken at
other times.

\begin{figure}
\centerline{\includegraphics[width=0.52\textwidth]{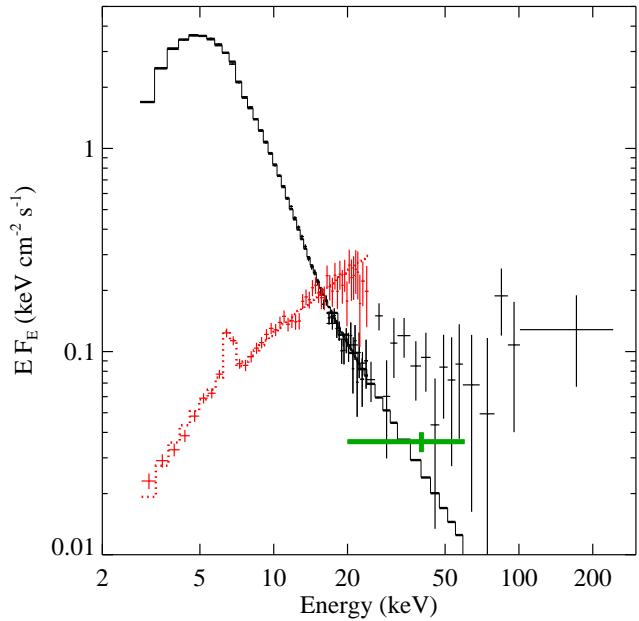}}
\caption{Energy spectra and flux measurements 
demonstrating that it is likely that the high energy excess
seen for some of the {\em RXTE} observations comes from the
nearby source IGR~J16320--4751.  The black spectrum (thin, 
solid histogram) demonstrates the presence of a high energy 
excess in the {\em RXTE} observation 70417-01-07-02.  
The thick, solid green point (20--60 keV) represents the 
IGR~J16320--4751 flux measured by {\em INTEGRAL} in 2003 
February.  The red PCA spectrum (dotted histogram) comes 
from an {\em RXTE} observation pointed at 4U~1630--47 during 
a time in early 2000 (Observation ID 40112-01-31-00) when 
4U~1630--47 was in quiescence.\label{fig:spectra_a}}
\end{figure}

Our 2003 February {\em INTEGRAL} observation provides the cleanest
measurement of the high energy flux from IGR J16320--4751.  From
the \cite{foschini04} analysis of the ISGRI spectrum, the peak 
20--60~keV flux during this observation is 
$2\times 10^{-10}$ erg~cm$^{-2}$~s$^{-1}$, and the energy spectrum
is consistent with a power-law with $\Gamma$ between 2.6 and 3.1.
In Figure~\ref{fig:spectra_a}, we plot the {\em RXTE} spectrum
for Observation ID 70417-01-07-02 and the 20--60~keV flux 
measured by {\em INTEGRAL} $\sim$135 days later.  We reduced 
the {\em INTEGRAL} flux appropriately to account for the {\em RXTE}
collimator response at the location of IGR J16320--4751.  Although 
the measured {\em INTEGRAL} flux at the time of the {\em INTEGRAL} 
observation is a factor of $\sim$2--3 too low to explain the strong 
residuals, long-term variability at higher levels than this has
been reported from IGR J16320--4751 \citep{rodriguez03}, making
it probable that the source gets bright enough to explain the
strong residuals at times.

Further evidence that IGR J16320--4751 is sometimes bright enough
to explain the strong residuals comes from {\em RXTE} observations
made of 4U~1630--47 between 1999 December 25 and 2000 January 14.
These observations were made when 4U~1630--47 was in quiescence
in an effort to observe 4U~1630--47 as it turned-on; however, 
it did not turn on during this sequence of 36 very short
(typically 500~s) observations.  During this sequence of 
observations (Observation IDs 40112-01-01-00 to 40112-01-36-00)
the PCA count rate varied significantly between 8 and 26 
c/s/PCU.  Although some of the detected emission is probably
Galactic Ridge emission, the high level of variability indicates
that a compact source is producing much of the flux, and 
IGR J16320--4751 is the most likely candidate.  In 
Figure~\ref{fig:spectra_a}, we show the PCA spectrum from the
observation with the highest count rate, which was made on 
2000 January 11 (Observation ID 40112-01-31-00).  We note that 
no collimator response correction is necessary because the 
pointing position is approximately the same for this observation 
and for 70417-01-07-02.  Another technical note is that although
we attempted to also use HEXTE, with such short observations in
a crowded field, this was not straightforward.  
Figure~\ref{fig:spectra_a} indicates that the source was very 
hard $\Gamma\sim 1.3$, and the extension of this spectrum is more 
than bright enough to explain the high energy residuals, even if 
the spectrum breaks above 20~keV.  In summary, this analysis 
provides strong evidence that IGR J16320--4751 is the cause of 
the high energy residuals.

\begin{center}
{\bf\normalsize Appendix B}
\end{center}

In addition to the four observations discussed in Appendix A, 
there are seven other observations for which it is likely that 
the high energy emission is dominated by a source other than 
4U~1630--47.  The spectral parameters for these seven 
observations are shown in Figure~\ref{fig:parameters}, and 
these are the extremely hard observations with $\Gamma$ in 
the 0.6--1.7 range.  As discussed in \S$5$ above, four of
the seven observations (70113-01-20-00, 70113-01-21-00, 
70113-01-25-00, and 70113-01-28-00) occurred during a time 
period when IGR J16358--4726 was active according to reports
from {\em INTEGRAL}, {\em Chandra} and other {\em RXTE} 
observations \citep{revnivtsev03a,revnivtsev03b,patel04}.
As was the case for IGR J16320--4751, {\em INTEGRAL} provides
the cleanest measurement of the high energy flux due to its
imaging capabilities, and \cite{revnivtsev03b} report a 
15--40 keV flux of 50 mCrab ($\sim$$2\times 10^{-9}$ 
erg~cm$^{-1}$~s$^{-1}$) and a 40--100 keV flux of
20 mCrab ($\sim$$3\times 10^{-10}$ erg~cm$^{-1}$~s$^{-1}$).
on MJD 52,727.4.  Our observation 70113-01-25-00 occurred
on the same day (MJD 52,727.633), and, after correcting
the {\em RXTE} collimator response for the position of 
IGR J16358--4726, which is $0^{\circ}.61$ from the center
of the ``Offset \#1'' FOV, we measure 15--40 and 40--100
keV fluxes of $7.5\times 10^{-10}$ and $2.5\times 10^{-10}$ 
erg~cm$^{-1}$~s$^{-1}$, respectively.  The 15--40~keV
flux is a factor of 2.7 lower than the {\em INTEGRAL}
measurement, and the 40--100~keV is nearly the same in
the two cases.  Although the 15--40~keV flux measured by
{\em RXTE} is somewhat low, the difference is consistent
with a report by \cite{revnivtsev03b} that the flux 
varies by a factor of $\sim$2 on a time scale of hours.
More importantly, the flux comparison confirms our suspicion
that the 4U~1630--47 spectrum is contaminated by the flux 
from IGR J16358--4726 at high energies.

\begin{figure}
\centerline{\includegraphics[width=0.52\textwidth]{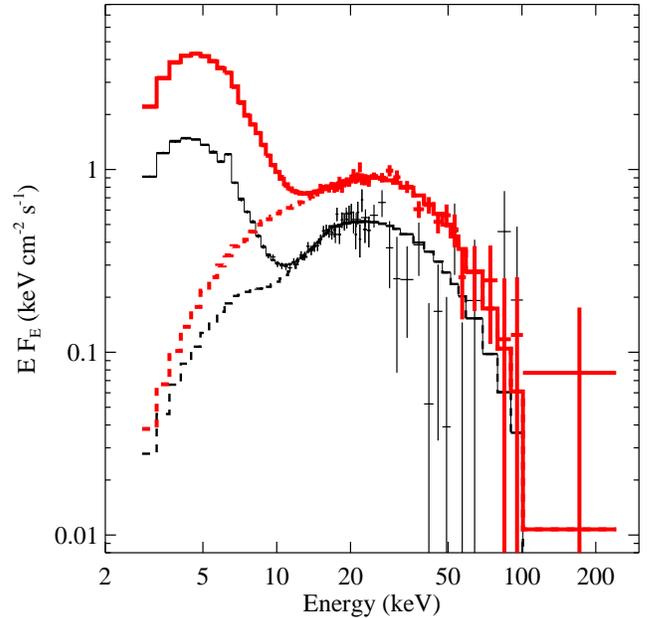}}
\caption{{\em RXTE} spectra demonstrating
the high energy contribution from other sources.  The
soft component comes from 4U~1630--47, but the emission 
above $\sim$15 keV is dominated by another source.  The 
black {\em RXTE} spectrum (thin histogram) from Observation 
ID 70113-01-25-00 was taken on the same day as an 
{\em INTEGRAL} observation of the region, and we argue
in the text that it is very likely that the high energy
flux comes from IGR J16358--4726.  The red {\em RXTE} 
spectrum (thick histogram) comes from much later 
(Observation ID 90410-01-01-01 taken on MJD 53,114.410), 
but the two spectra are very similar, suggesting that 
they may be from the same source.\label{fig:spectra_b}}
\end{figure}

Although it is clear that we are seeing emission from
IGR J16358--4726 in four of the seven observations with
low values of $\Gamma$, it is less clear whether the
high energy emission for the other three observations:
80420-01-15-00 made on MJD 53,002.863; 80117-01-24-00
made on MJD 53,076.664; 90410-01-01-01 made on 
MJD 53,114.410; is also dominated by IGR J16358--4726
because no {\em INTEGRAL} detections of IGR J16358--4726 
were reported and also because IGR J16320--4751 was in
the {\em RXTE} FOV for these three observations.  In 
Figure~\ref{fig:spectra_b}, we compare the PCA plus HEXTE 
spectrum from 90410-01-01-01 to that of 70113-01-25-00, 
where we know the emission is from IGR J16358--4726, and 
the spectra are remarkably similar.  After correcting for 
the {\em RXTE} collimator response, the 15--40 and 
40--100~keV fluxes for 90410-01-01-01 are about a factor 
of 1.8 higher than for 70113-01-25-00.  Even though
this may be considered an argument in favor of the 
emission being from IGR J16358--4726, IGR J16320--4751
is also known to produce very hard ($\Gamma < 1.0$)
spectra at times \citep{tomsick03_iauc}.  

\clearpage



\clearpage

\begin{table}
\caption{{\em RXTE} Observations\label{tab:observations}}
\begin{minipage}{\linewidth}
\footnotesize
\begin{tabular}{cccc} \hline \hline
Proposal & Number of     &  Range of Exposure & Mean Exposure\\
ID       & Observations  &  Times (s)         & Time (s)\\ \hline \hline
P70417   &  53           &  96 -- 7504        & 1745\\
P70113   &  55           &  288 -- 11216      & 2756\\
P80117   &  53           &  96 -- 7376        & 1849\\
P80417   &   1           &  1072 -- 1072      & 1072\\
P80420   &  55           &  320 -- 2320       & 1144\\
P90128   &  54           &  256 -- 3392       & 1635\\
P90410   &  47           &  464 -- 3264       & 1775\\ \hline
\end{tabular}
\end{minipage}
\end{table}

\begin{table}
\caption{Spectral Parameters\label{tab:spectra}}
\begin{minipage}{\linewidth}
\scriptsize
\begin{tabular}{ccccccccc} \hline \hline
Observation & MJD\footnote{Modified Julian Date.} & $N_{\rm H}$\footnote{Hydrogen column density in units of $10^{22}$ cm$^{-2}$.  For all the parameters, 90\% confidence errors are given.} & $kT_{\rm in}$\footnote{The inner disk temperature measured with the disk-blackbody (``diskbb'') model.} & $N_{\rm diskbb}$\footnote{The normalization for the disk-blackbody component;  $N_{\rm diskbb} = (R_{\rm in}/d_{10})^2\cos i$, where $R_{\rm in}$ is the disk inner radius in units of km, $d_{10}$ is the distance to the source in units of 10 kpc, and $i$ is the disk inclination.} & $\Gamma$\footnote{Power-law photon index.} & $PLR$\footnote{Ratio of the unabsorbed 2--20 keV flux in the power-law component to the total 2--20 keV flux.} & Flux\footnote{Unabsorbed 3--200 keV flux in erg cm$^{-2}$ s$^{-1}$.} & Spectral\footnote{SPL, TD, IS, and Hard state label the spectral state of the observation.  The observation labeled ``Flaring'' is an IS observation that exhibits the flaring behavior.}\\
ID & & (cm$^{-2}$) & (keV) & & & & & State\\ \hline \hline
70417-01-09-00 & 52,636.852 & $11.03^{+0.28}_{-0.21}$ & $3.22^{+0.10}_{-0.05}$ & $4.2^{+0.4}_{-0.8}$ & $2.71^{+0.03}_{-0.02}$ & $0.835^{+0.014}_{-0.009}$ & $3.90\times 10^{-8}$ & SPL\\
70113-01-43-00 & 52,780.746 & $9.76^{+0.69}_{-0.36}$ & $1.58^{+0.02}_{-0.03}$ & $230\pm 20$ & $2.50^{+0.22}_{-0.20}$ & $0.34^{+0.10}_{-0.05}$ & $2.50\times 10^{-8}$ & Flaring\\
70417-01-06-00 & 52,558.188 & $9.67^{+1.13}_{-0.17}$ & $1.39^{+0.01}_{-0.02}$ & $278^{+11}_{-20}$ & $3.10^{+0.50}_{-0.10}$ & $0.21^{+0.16}_{-0.04}$ & $1.16\times 10^{-8}$ & TD\\
70113-02-04-00 & 52,675.941 & $7.13^{+1.07}_{-0.68}$ & $0.84^{+0.05}_{-0.06}$ & $437^{+308}_{-132}$ & $2.24^{+0.06}_{-0.04}$ & $0.40\pm 0.01$ & $2.35\times 10^{-9}$ & IS\\
80117-01-21-01 & 53,068.789 & $<$8.9 & $0.50^{+0.15}_{-0.05}$ & $1246^{+4254}_{-846}$ & $1.92^{+0.05}_{-0.06}$ & $0.80^{+0.02}_{-0.14}$ & $1.75\times 10^{-9}$ & Hard\\ \hline 
\end{tabular}
\end{minipage}
\end{table}

\begin{table}
\caption{Timing Parameters\label{tab:power}}
\begin{minipage}{\linewidth}
\scriptsize
\begin{tabular}{ccccccccc} \hline \hline
Observation & Spectral\footnote{SPL, TD, IS, and Hard state label the spectral state of the observation.  The observation labeled ``Flaring'' is an IS observation that exhibits the flaring behavior.} & Model\footnote{Components used in modeling the
power spectra, where PL = power-law, BL = band-limited, and QPO = 
quasi-periodic oscillation.} & Continuum & $\nu_{\rm QPO}$ & $Q$\footnote{The quality factor of the QPO, defined as $\nu_{\rm QPO}$ divided by the QPO's FWHM.} & QPO & $A_{\rm pp}$\footnote{The peak-to-peak amplitude of the noise (see the text for a precise definition.)}\\
ID & State & & rms (0.1--10 Hz) & (Hz) & & rms\\ \hline \hline
70417-01-09-00 & SPL & PL+QPO & 1.8\%$\pm$0.2\% & $12.15\pm 0.15$ & $4.4\pm 0.7$ & 1.30\%$\pm$0.07\% & $0.25\pm 0.01$\\
70113-01-43-00 & Flaring & PL & 6.1\%$\pm$0.5\% & -- & -- & -- & $0.86\pm 0.03$\\
70417-01-06-00 & TD & PL & 5.8\%$\pm$0.7\% & -- & -- & -- & $0.26\pm 0.01$\\
70113-02-04-00 & IS & PL+BL & 3.5\%$\pm$1.3\% & -- & -- & -- & $0.31\pm 0.04$\\
80117-01-21-01 & Hard & BL+QPO & 9.6\%$\pm$0.4\% & $5.13\pm 0.01$ & $21\pm 5$ & 5.4\%$\pm$ 0.3\% & $0.20\pm 0.05$\\ \hline
\end{tabular}
\end{minipage}
\end{table}

\end{document}